\documentclass[12pt]{article}
\usepackage{amsmath,amssymb}
\newcommand{\beq}{\begin{equation}}
\newcommand{\eeq}{\end{equation}}
\newcommand{\ba}{\begin{array}}
\newcommand{\ea}{\end{array}}
\newcommand{\bea}{\begin{eqnarray}}
\newcommand{\eea}{\end{eqnarray}}
\newcommand{\bean}{\begin{eqnarray*}}
\newcommand{\eean}{\end{eqnarray*}}

\newtheorem{theorem}{Theorem}[section]
\newtheorem{prop}[theorem]{Proposition}

\newtheorem{exe}[theorem]{Exercise}
\newtheorem{defi}[theorem]{Definition}
\newtheorem{remark}[theorem]{Remark}

\newenvironment{exer}{\begin{exe} \rm}{\end{exe}}
\newtheorem{proof}{Proof.}
\newenvironment{pf}{\begin{proof} \rm}{\hfill$\square$ \end{proof}}
\newcommand{\CH}{{\cal H}}

\newcommand{\CD}{{\cal D}}
\newcommand{\CW}{{\cal W}}

\newcommand{\CS}{{\cal S}}

\newcommand{\CF}{{\cal F}}
\newcommand{\CM}{{\cal M}}

\newcommand{\CL}{{\cal L}}

\makeatletter
\@addtoreset{equation}{section}

\makeatother
\newcommand{\Hh}{{\mathsf H}}
\newcommand{\Vv}{{\mathsf V}}
\newcommand{\Uu}{{\mathsf U}}
\newcommand{\Ww}{{\mathsf W}}
\newcommand{\Ff}{{\mathsf F}}
\newcommand{\Tt}{{\mathsf T}}

\newcommand{\RR}{{\mathbb R}}

\newcommand{\NN}{{\mathbb N}}

         \def\Ga{\Gamma}
\def\be{\beta}
\def\al{\alpha}

\def\la{\lambda}        \def\La{\Lambda}

\newcommand{\faa}[3]{Funct. Anal. Appl. {\bf #1} (#2), #3}

\newcommand{\lmp}[3]{Lett. Math. Phys. {\bf #1} (#2), #3}

\newcommand{\jmp}[3]{Jour. Math. Phys. {\bf #1} (#2), #3}
\newcommand{\rref}[1]{(\ref{#1})} 
\def\binomial#1#2{{#1\choose #2}}

\def\dsl{\displaystyle}

\newcommand{\del}{{\partial}}

\def\parpo#1#2{\{#1,#2\}}
\def\mat2#1#2#3#4{{\left(\begin{array}{cc}#1 & #2\\ #3 & #4 \end{array}\right)}}
\def\mats2#1#2#3#4{{\left(\begin{array}{cc}#1 & #2\vspace{2truemm} \\ #3 & #4 
\end{array}\right)}}
\def\vec2#1#2{{\left[\begin{array}{c}
#1 \\ #2 \end{array}\right]}}
\def\Bdpt#1#2{\displaystyle{\frac{\partial #1}{\partial t_{#2}}}}
\def\dpt#1#2{{\frac{\partial #1}{\partial t_{#2}}}}
\def\ddd#1#2{\displaystyle{\frac{\partial #1}{\partial #2}}}
\def\Hdd#1{\displaystyle{\frac{\partial H}{\partial h_{#1}}}}
\def\fddd#1#2{\displaystyle{\frac{\delta #1}{\delta #2}}}
\newcommand{\DT}{\frac{d}{dt}}
\newcommand{\Ha}[1]{H^{(#1)}}
\newcommand{\h}[1]{h^{(#1)}}
\newcommand{\Wa}[1]{W^{(#1)}}
\newcommand{\ala}[1]{\al^{(#1)}}
\newcommand{\bb}[1]{\be^{(#1)}}
\newcommand{\Lax}[1]{L^{(#1)}}
\newcommand{\eqcon}[1]{{\buildrel{(#1)}\over{=}}}

\newcommand{\ha}[1]{h^{(#1)}}
\def\Nij{Nijenhuis}
\def\HJ{Hamilton--Jacobi\ }

\def\alg{{\mathfrak g}}
\def\algd{{\mathfrak g}^*}

\newcommand{\fraksl}{{\mathfrak s}{\mathfrak l}}
\def\ger{hierarch}
\def\var{manifold}

\def\bih{bi-Ham\-il\-tonian}
\def\varb{\bih\ \var}
\def\ham{Hamiltonian}
\def\vefi{vector field}

\def\parp{Poisson bracket}
\def\ger{hierarch}

\def\parpu{{\parpo{\cdot}{\cdot}}}

\begin{document}
\begin{flushright}
Ref. SISSA 135/1999/FM
\end{flushright}
\begin{center}
{\huge \bf
The method of Poisson pairs in the theory of nonlinear 
PDEs\footnote{Work partially supported by
the M.U.R.S.T. and the G.N.F.M. of the Italian C.N.R.\\
Lectures given by F. Magri at the 1999 CIME course
``Direct and Inverse Methods in Solving Nonlinear Evolution
Equations '' Cetraro, (Italy)  September 1999.       } }
\end{center}
\vspace{0.8truecm}
\makeatletter
\begin{center}
{\large
Franco Magri${}^a$,
Gregorio Falqui${}^b$,
 and
Marco Pedroni${}^c$}\\ \bigskip
${}^a$ Dipartimento di Matematica e Applicazioni\\ 
Universit\`a  di Milano--Bicocca\\
Via degli Arcimboldi 8, I-20126 Milano, Italy\\
E--mail: magri@vmimat.mat.unimi.it\\
${}^b$ SISSA, Via Beirut 2/4, I-34014 Trieste, Italy\\
E--mail: falqui@sissa.it\\
${}^c$ Dipartimento di Matematica, Universit\`a di Genova\\
Via Dodecaneso 35, I-16146 Genova, Italy\\
E--mail: pedroni@dima.unige.it
\end{center}
\makeatother
\vspace{0.1truecm}
\begin{abstract}
\noindent
The aim of these lectures is to show that the methods of classical 
Hamiltonian mechanics  can be profitably used to solve certain classes of 
nonlinear partial differential equations. 
The prototype of these equations is the 
well-known Korteweg--de Vries (KdV) equation.\\
In these lectures we touch the following subjects:
\begin{enumerate}
\item[i)] the birth and the role of the method of Poisson pairs inside the
  theory of the KdV equation;
\item[ii)] the theoretical basis of the method of Poisson pairs;
\item[iii)] the Gel'fand--Zakharevich theory of integrable systems on \varb s;
\item[iv)] the Hamiltonian interpretation of the Sato picture of the KdV flows 
  and of its linearization on an infinite--dimensional Grassmannian manifold.
\item[v)] the reduction technique(s) and its use to construct classes of
  solutions;
\item[vi)] the role of the technique of separation of variables in the study
  of the reduced systems;
\item[vii)] some relations intertwining the method of Poisson pairs with the
  method of Lax pairs.
\end{enumerate}
\end{abstract}
\tableofcontents
\newpage
\section{Introduction: The tensorial approach and the birth of the 
method of Poisson
  pairs}\label{lect:1} This lecture is an introduction to the Hamiltonian
analysis of PDEs form an ``experimental'' point of view. This means that we
are more concerned in unveiling the spirit of the method than in working out
the theoretical details. Therefore the style of the exposition will be
informal, and proofs will be mainly omitted. We shall follow, step by step,
the birth and the evolution of the Hamiltonian analysis of the KdV equation
\begin{equation}\label{eq:1.1}
u_t=\frac{1}{4}(u_{xxx}-6uu_{x})\ ,
\end{equation}
from its ``infancy'' to the final representation of the KdV flow as a linear
flow on an infinite-dimensional Grassmannian due to Sato~\cite{SS82}. The 
route is long and demanding. Therefore the exposition is divided in two parts, 
to be carried out in this and in the fourth lecture (see
Section~\ref{lect:4}). Here our primary aim is to show the birth of the method 
of Poisson pairs. It is reached by means of a suitable use of the well-known
methods of tensor analysis. We proceed in three steps. First, by using the
transformation laws of vector fields, we construct the Miura map and the so
called {\em modified} KdV equation (mKdV). This result leads quite simply to
the theory of (elementary) Darboux transformations and to the concept of
Poisson pair. Indeed, a peculiarity of mKdV is to possess an elementary
Hamiltonian structure. By means of the transformation law of Poisson bivectors, 
we are then able to transplant this structure to the KdV equation, unraveling 
its ``\bih\ structure''.
This structure can be used in turn to define the concept of {\em Lenard
  chain} and to plunge the KdV equation into the ``KdV hierarchy''. This step
is rather important from the point of view of finding classes of solutions
to the KdV equation. Indeed the hierarchy is a powerful instrument to
construct finite-dimensional invariant submanifolds of the  equation and,
therefore, finite-dimensional reductions of the KdV equation. The study of
this process of restriction and of its use to construct solutions will be one
of the two leading themes of these lectures. It is intimately related to the
theory of separation of variables dealt with in the last two lectures. The
second theme is that of the linearization of the full KdV flow on the 
infinite-dimensional Sato Grassmannian. The starting point of this process is
surprisingly simple, and once again based on a simple procedure of tensor
calculus. By means of the transformation laws of one--forms, we pull back the
KdV hierarchy from its phase space onto the phase space of the mKdV
equation. In this way we obtain the ``mKdV hierarchy''. In the fourth lecture
we shall show that this hierarchy can be written as a flow on an
infinite--dimensional Grassmann manifold, and that this flow can be linearized 
by means of a (generalized) Darboux transformation.
\subsection{The Miura map and the KdV equation}\label{lect:11}
As an effective way of probing the properties of equation \rref{eq:1.1} we 
follow
the tensorial approach. Accordingly, we regard equation \rref{eq:1.1} as the 
{\em
  definition} of a vector field
\begin{equation}\label{eq:1.2}
u_t=X(u,u_x,u_{xx},u_{xxx})
\end{equation}
on a suitable function space, and we investigate how it transforms under a
point transformation in this space. Since our ``coordinate''
$u$ is a function and not simply a number, we are allowed to consider 
transformations of coordinates depending also on the derivatives of the 
new coordinate function of the type\footnote{For further details on these kind 
  of transformations, see~\cite{Di-B91}.}
\begin{equation}\label{eq:1.3}
  u=\Phi(h,h_x)\ .
\end{equation}
We ask whether there exists a transformed vector field
\begin{equation}\label{eq:1.4}
  h_t=Y(h,h_x,h_{xx},h_{xxx})
\end{equation}
related to the KdV equation according to the transformation law for vector
fields, 
\begin{equation}\label{eq:1.5}
  X(\Phi(h))=\Phi_h^\prime\cdot Y(h),
\end{equation}
where $\Phi_h^\prime$ is the (Fr\'echet) derivative of the operator $\Phi$
defining the transformation. This condition gives rise to a (generally
speaking) over-determined system of partial differential equations on the
unknown functions $\Phi(h,h_x)$ and $Y(h,h_x,h_{xx},h_{xxx}).$
In the specific example the over-determined system can be solved. Apart form
the trivial solution $u=h_x$, we find the {\em Miura transformation\/} 
\cite{MGK68,Ku81},
\begin{equation}\label{eq:1.6}
  u=h_x+h^2-\lambda\ ,
\end{equation}
depending on an arbitrary parameter $\la$. The transformed equation is
the {\em modified} KdV equation:
\begin{equation}
  \label{eq:1.7}
  h_t=\frac14(h_{xxx}-6h^2h_x+6\la h_x)\ .
\end{equation}
\begin{exer}
Work out in detail the transformation law~\rref{eq:1.5}, checking that
$X$, $\Phi$, and $Y$, defined respectively by equations \rref{eq:1.1},
\rref{eq:1.6} and~\rref{eq:1.7} do satisfy equation \rref{eq:1.5}.
\hfill $\square$\end{exer}
The above result is plenty of consequences. The first one is a simple method
for constructing solutions of the KdV equation. It is called the method of
(elementary) Darboux transformations~\cite{Matveev-Salle}. It rests on the
remark that the mKdV equation~\rref{eq:1.7} admits the discrete symmetry
\begin{equation}
  \label{eq:1.8}
  h\mapsto h^\prime=-h\ .
\end{equation}
Let us exploit this property to construct the well-known one--soliton solution 
of the KdV equation. We notice that the point $u=0$ is a very simple
(singular) invariant submanifold of the KdV equation. Its inverse image under
the Miura transformation is the 1--dimensional submanifold $S_1$ formed by the
solutions of the special Riccati equation
\begin{equation}
  \label{eq:ric0}
  h_x+h^2=\la\ .
\end{equation}
This submanifold, in its turn, is invariant with respect to
equation \rref{eq:1.7}.  A straightforward computation shows that, on this
submanifold,
\begin{equation}
  \label{eq1.9}
  \frac14(h_{xxx}-6h^2h_x+6\la h_x)=\la h_x\ .
\end{equation}
Therefore, on $S_1$ the mKdV equation takes the simple form $h_t=\la h_x.$
Solving the first order system formed by this equation  and the the 
Riccati equation $h_x+h^2=\la$, and setting $\la=z^2$, we find the general
solution
\begin{equation}
  \label{eq:1.10}
  h(x,t)=z\, \tanh(zx+z^3t+c)
\end{equation}
of the mKdV equation on the invariant submanifold $S_1.$ At this point we use
the symmetry property and the Miura map. By the symmetry
property~\rref{eq:1.8} the function $-h(x,t)$ is a new solution of the
modified equation, and by the Miura map the function
\begin{equation}
  \label{eq:1.11}
  u^\prime(x,t)=-h_x+h^2-z^2=2z^2\mbox{sech}^2(zx+z^3t+c)
\end{equation}
is a new solution of the KdV equation. It is called the {\em one soliton} 
solution\footnote{For a very nice account of  the origin and  of the
 properties  of the KdV equation and of other soliton equations and their
 solutions, see, e.g., \cite{Ne-B85}.}.  It can also be interpreted in terms of
invariant submanifolds. To this end, we have to notice that the Miura map~\rref 
{eq:1.6}
transforms the invariant submanifold $S_1$ of the modified equation into the
submanifold formed by the solutions of the first order differential equation
\begin{equation}
  \label{eq:1.12}
  \frac{1}{2}\left(-\frac12 u_x^2+u^3\right)+\la u^2=0\ .
\end{equation}
As one can easily check, this set is preserved by the KdV
equation, and therefore  it is an invariant one--dimensional submanifold of
the KdV equation, built up from the singular manifold $u=0$. On this
submanifold,  the KdV equation takes the simple form $u_t=\la u_x$, and the
flow can be integrated to recover the solution~\rref{eq:1.11}.

This example clearly shows that the Darboux transformations are a mechanism to
build invariant submanifolds of the KdV equation. Some of these submanifolds
will be examined in great detail in the present lectures. The purpose is to
show that the reduced equations on these submanifolds are classical
Hamiltonian vector fields whose associated Hamilton--Jacobi equations can be
solved by separation of variables. In this way, we hope, the interest of the
Hamiltonian analysis of the KdV equations can better emerge.
\subsection{Poisson pairs and the KdV hierarchy}\label{lect:12}
We shall now examine a more deep and far reaching property of the Miura
map. It is connected with the concept of Hamiltonian vector field. From
Analytical Mechanics, we know that the Hamiltonian vector fields are the
images of exact one--forms through a suitable linear map, associated with a
so--called Poisson bivector. We shall formally define these notions in the
next lecture. These definition can be easily extended to vector fields on
infinite--dimensional manifolds. Let us give an example, by showing that the
mKdV equation is a Hamiltonian vector field.  This requires a series of three
consecutive remarks. First we notice that equation~\rref{eq:1.7} can be
factorized as
\begin{equation}
  \label{eq:1.13}
  h_t=\left[\frac12 \del_x\right]\cdot\left[ \frac12 h_xx-\frac32 h^3+3\la 
h\right]\ .
\end{equation}
 Then, we notice that the linear operator in the first bracket, 
$\frac12\del_x$, is 
 a constant skewsymmetric operator which we can recognize as a Poisson
 bivector. Finally, we notice that in the differential polynomial appearing in 
 the second bracket in the right hand side of equation \rref{eq:1.13},
we can easily recognize an exact one--form. Indeed,
\begin{equation}
  \label{eq:1.14}
  \int
\left(\frac12 h_{xx}-\frac32 h^3+3\la h\right)\dot h\, dx= 
\DT \int
\left(
  -\frac12 h_x^2-\frac38 h^4+\frac32 \la h^2\right)\, dx\
\end{equation}
for any tangent vector $\dot h$. 
These statements are true under suitable boundary conditions, as explained in,
e.g.,~\cite{Di-B91}.
Here and in the rest of these lectures we will tacitly use periodic boundary
conditions.

The Hamiltonian character of the mKdV equation is obviously independent of the 
existence of the Miura map. However, this map finely combines this property
from the point of view of tensor analysis. Let us recall that a Poisson
bivector is a skewsymmetric linear map from the cotangent to the tangent spaces 
satisfying a 
suitable differential condition (see Lecture~\ref{lect:2}).
It obeys the transformation law
\begin{equation}
  \label{eq:1.15}
  Q_{\Phi(h)}=\Phi^\prime_h P_h \Phi^{\prime\,*}_h
\end{equation}
under a change of coordinates (or a map between two different manifolds). In
this formula the point transformation is denoted (in operator form) by
$u=\Phi(h)$. The operators $\Phi^\prime_h$ and $\Phi^{\prime\, *}_h$ are the
Fr\'echet derivative of $\Phi$ and its adjoint operator.  The symbols $P_h$
and $Q_u$ denote the Poisson bivectors in the space of the functions  $h$ and
$u$, respectively. 
Since the Miura map $u=h_x+h^2-\la$ is not invertible, it
is rather nontrivial that there exists a Poisson bivector $Q_u$, on the
phase space of the KdV equation, which is $\Phi$--related (in the sense of
equation \rref{eq:1.15}) to the Poisson bivector $P_h=\frac12 \del_x$
associated with the modified equation. Surprisingly, this is the case. One can
check that the operator $Q_u$ is defined by
\begin{equation}
  \label{eq:1.16}
  Q_u=-\frac12\del_{xxx}+2(u+\la)\del_x+u_x\ .
\end{equation}
\begin{exer} 
Verify the above claim by computing the product (in the
  appropriate order) of the operators $\Phi^\prime_h=\del_x+h$,
  $P_h={\frac12 \del_x}$, and $\Phi^{\prime\, *}_h=-\del_x+h$, and
  by expressing the results in term of $u=h_x+h^2-\la$.
\hfill $\square$\end{exer}
This exercise shows that the Miura map is a peculiar Poisson map. Since it
depends on the parameter $\la$, the final result is that the phase space of
the KdV equation is endowed with a one--parameter family of Poisson bivectors,
\begin{equation}\label{eq:1.17}
  Q_\la=Q_1-\la Q_0\ ,
\end{equation}
which we call a Poisson pencil. The operators $(Q_1,Q_0)$ defining the pencil
are said to form a {\em Poisson pair}, a concept to be systematically
investigated in the next lecture.
 
These operators enjoy a number of interesting properties, and define new
geometrical structures associated with the equation. One of the simplest but
far--reaching is the concept of {\em Lenard chain}. The idea is to use
the pair of bivectors to define a recursion relation on one--forms:
\begin{equation}
  \label{eq:1.18}
  Q_0\alpha_{j+1}=Q_1\alpha_{j}\ .
\end{equation}
In the applications a certain care must be taken in dealing with this
recursion relation, since it does not define uniquely the forms $\al_j$ (the
operator $Q_0$ is seldom invertible). Furthermore, it is still less apparent
that it can be solved in the class of {\em exact} one--forms. However, in the
KdV case we {\em bonafide\/}  proceed and we find
\begin{equation}
  \label{eq:1.19}
  \begin{split}
\al_0&=1\\
\al_1&=-\frac12 u\\
\al_2&=\frac18(-u_{xx}+3u^2)\\
\al_3&=\frac{1}{32}(-10 u^{3}+10\,u u_{xx}-u_{xxxx}+5u_x^{2} )\\
\end{split}
\end{equation}
as first terms of the recurrence. The next step is to consider the associated
vector fields (the meaning of numbering them with {odd} integers will be
explained in Lecture 4):
\begin{equation}
  \label{eq:1.20}
  \begin{split}
\Bdpt{u}{1}&=Q_1\al_0=Q_0\al_1=u_x\\
\Bdpt{u}{3}&=Q_1\al_1=Q_0\al_2=\frac{1}{4}(u_{xxx}-6 u u_x)\\
\Bdpt{u}{5}&=Q_1\al_2=Q_0\al_3=\frac{1}{16}(u_{xxxxx}-10 u u_{xxx}-20 
u_xu_{xx}+30u^2u_x)\ .\\
\end{split}
\end{equation}
They are the first members of the KdV hierarchy. In the fourth lecture,
we shall show that it is a special instance of a general concept, the
 Gel'fand--Zakharevich hierarchy associated with any Poisson pencil of a
suitable class.
\subsection{Invariant submanifolds and reduced equations}\label{lect:13}
The introduction of the KdV {hierarchy} has important consequences on the
problem of constructing solutions of the KdV {equation}. The hierarchy is
indeed a basic supply of invariant submanifolds of the KdV equation. This is
due to the property of the vector fields of the hierarchy to commute among
themselves. From this property it follows that the set of singular point of
any linear combination (with constant coefficients)
of the vector fields of the hierarchy is a
finite--dimensional invariant submanifold of the KdV flow. These submanifold
can be usefully exploited to construct classes of solutions of this
equation.

As a first example of this technique we consider the submanifold
defined by the condition
\begin{equation}
  \label{eq:1.21}
  \Bdpt{u}{3}=\la  \Bdpt{u}{1}\>,
\end{equation}
that is, the submanifold where the second  vector field of the \ger y is a 
constant multiple 
of the first one. It is formed by the solutions of the third order
differential equation
\begin{equation}
  \label{eq:1.22}
  \frac{1}{4}(u_{xxx}-6 u u_x)-\la u_x=0\ .
\end{equation}
Therefore it is a three dimensional manifold, which we denote by $M_3$. We can
use as coordinates on $M_3$ the {\em values} of the function $u$ and its
derivatives $u_x$ and $u_{xx}$ at {\em any} point $x_0$.
To avoid
cumbersome notations, we will continue to denote these three {\em numbers} 
with the same symbols, $u,u_x,u_{xx}$, but the reader should be
aware of this subtlety. To perform the reduction of the first equation of the
hierarchy~\rref{eq:1.20} on $M_3$, we consider the first two differential
consequences of the equation $\Bdpt{u}{1}=u_x$ and we use the
constraint~\rref{eq:1.21} to eliminate the third order derivative.
We obtain the system
\begin{equation}
  \label{eq:1.23}
  \Bdpt{u}{1}=u_x,\quad  \Bdpt{u_x}{1}=u_{xx},\quad  \Bdpt{u_{xx}}{1}=6uu_x+4\la 
  u_x\ .
\end{equation}
We call $X_1$ the vector field defined by these equations on $M_3$. It shares
many of the properties of the KdV equation. For instance, it is related to a
Poisson pair. The simplest way to display this property is to remark that
$X_1$ possesses two integrals of motion,
\begin{equation}
  \label{eq:1.24}
  \begin{split}
H_0&=u_{xx}-3u^2-4\la u\\
H_1&=-\frac12 u_x^2+u^3+2\la u^2+u H_0\ .\\
\end{split}
\end{equation}
 Then we notice that on $M_3$ there exists a unique Poisson bracket
 ${\parpu}_0$ with the following two properties:
 \begin{enumerate}
 \item[i)] The function $H_0$ is a Casimir function, that is, 
${\parpo{F}{H_0}}_0=0$ for every smooth
   function $F$ on $M_3$.   
\item[ii)] $X_1$ is the Hamiltonian vector field associated with the function 
$H_1$.
 \end{enumerate}
Such a Poisson bracket $\parpu_0$ is defined by the relations
\begin{equation}
  \label{eq:1.25}
  {\parpo{u}{u_x}}_0=-1,\quad {\parpo{u}{u_{xx}}}_0=0,\quad 
{\parpo{u_x}{u_{xx}}}_0=6u+4\la\ .
\end{equation}
Similarly, one can notice that on $M_3$ there exists a unique Poisson bracket 
${\parpu}_1$ with the following ``dual'' properties:
 \begin{enumerate}
 \item[i')] The function $H_1$ is a Casimir function, that is, 
${\parpo{F}{H_1}}_1=0$ for every smooth function $F$ on $M_3$.
\item[ii')] $X_1$ is the Hamiltonian vector field associated with the function 
$H_0$.
 \end{enumerate}
This second Poisson bracket $\parpu_1$ is defined by the relations
\begin{equation}
  \label{eq:1.26}
  {\parpo{u}{u_x}}_1=u,\quad {\parpo{u}{u_{xx}}}_1=u_x,\quad
  {\parpo{u_x}{u_{xx}}}_1=u_{xx}-u(6u+4\la)\ .
\end{equation}
\begin{exer} Verify the stated properties of the Poisson pair~\rref{eq:1.25}
  and~\rref{eq:1.26}.\\
  \null\hfill $\square$\end{exer}
We now exploit the previous remarks to understand the geometry of the flow
associated with $X_1$. First we use the Hamiltonian representation
\begin{equation}
  \label{eq:1.27}
  \frac{dF}{dt}=X_1(F)={\parpo{F}{H_1}}_0\ .
\end{equation}
It entails that the level surfaces of the Casimir function $H_0$ are
two--dimensional (symplectic) manifolds to which $X_1$ is tangent.  Let us pick 
up any of these symplectic leaves, for instance the one passing through the
origin $u=0, u_x=0, u_{xx}=0$. Let us call $S_2$ this leaf. The coordinates
$(u,u_x)$ are {\em canonical} coordinates on $S_2$. The level curves of the
Hamiltonian $H_1$ define a Lagrangian foliation of $S_2$. Our problem is to find
the flow of the vector field $X_1$ along these Lagrangian submanifold. We have 
already given the solution of this problem in the particular case of the
Lagrangian submanifold passing through the origin. This submanifold is the
one--dimensional invariant submanifold~\rref{eq:1.12} previously discussed in
connection with Darboux transformations. The relative flow is the one--soliton 
solution to KdV. To deal with the generic Lagrangian submanifolds on an equal
footing, it is useful to change strategy and to use the Hamilton--Jacobi
equation
\begin{equation}
  \label{eq:1.28}
  H_1(u,\frac{dW}{du})=E\ .
\end{equation}
In this rather simple example, there is almost nothing to say about this
equation (which is obviously {solvable}), and the second Poisson
bracket~\rref{eq:1.25} seems not to play any role in the theory.
 
This (wrong!) impression is promptly corrected by the study of a more
elaborated example. Let us consider the five--dimensional submanifold $M_5$ of 
the  singular points of the {third} vector field of the KdV \ger y. 
It is defined 
by the equation
\begin{equation}
  \label{eq:1.29}
  u_{xxxxx}-10 u u_{xxx}-20 u_xu_{xx}+30u^2u_x=0\ .
\end{equation}
On this submanifold we can consider the restrictions of the first {\em two}
vector fields of the same hierarchy. To compute the reduced equation we
proceed as before. We regard the Cauchy data $(u,u_x,u_{xx},u_{xxx},u_{xxxx})$ 
as coordinates on $M_5.$ then we compute the time derivatives
$ \Bdpt{u}{1},\>\Bdpt{u_x}{1},
\>\Bdpt{u_{xx}}{1},\>\Bdpt{u_{xxx}}{1},\>\Bdpt{u_{xxxx}}{1} $ 
by taking the differential consequences of $\Bdpt{u}{1}=u_x$, and by
using~\rref{eq:1.29} and its differential consequences as a constraint to
eliminate all the derivatives of degree higher than four. We obtain the
equations
\begin{equation}
\label{eq:1.30}  
\begin{split}
\Bdpt{u}{1}&=u_x\\
\Bdpt{u_x}{1}&=u_{xx}\\
\Bdpt{u_{xx}}{1}&=u_{xxx}\\
\Bdpt{u_{xxx}}{1}&=u_{xxxx}\\
\Bdpt{u_{xxxx}}{1} &=10 u u_{xxx}+20 u_xu_{xx}-30u^2u_x
\end{split}
\end{equation}
In the same way, for the reduction of the KdV equation, we get
\begin{equation}
\label{eq:1.31}  
\begin{split}
\Bdpt{u}{3}&=\frac14(u_{xxx}-6uu_x)\\
\Bdpt{u_x}{3}&=\frac14(u_{xxxx}-6uu_{xx}-6u_x^2\\
\Bdpt{u_{xx}}{3}&=\frac14(4uu_{xxxx}+2 u_xu_{xx}-30u^2 u_x)\\
\Bdpt{u_{xxx}}{3}&=\frac14(2 u_{xx}^{2}+6 u_x u_{xxx}+4u u_{xxxx}-
60 u u_x^{2}-30u^2u_{xx})\\
\Bdpt{u_{xxxx}}{3} &=\frac14( 10 u_{xx} u_{xxx}+
10u_x u_{xxxx} -120 u^{3}u_x-100 u u_x u_{xx}\\ &\qquad +10u^{2}u_{xxx}-
60u_x^{3})\ .
\end{split}
\end{equation}
\begin{exer} Verify the previous computations.
\hfill $\square$\end{exer}
To find the corresponding solutions of the KdV equation, regarded as a partial 
differential equation in $x$ and $t$, we have to:
\begin{enumerate}
\item Construct a common solution to the ordinary differential
  equations~\rref{eq:1.30} and~\rref{eq:1.31};
\item Consider the first component $u(t_1,t_3)$ of such a solution;
\item Set $t_1=x$ and $t_3=t$. 
\end{enumerate}
The function $u(x,t)$ obtained in this way is the solution we were looking for. 
What
makes this procedure worth of interest is that the ODEs 
\rref{eq:1.30}--\rref{eq:1.31} 
can be solved by means of a variety of methods. In
particular, they can be solved by means of the method of separation of
variables\footnote{The fact that the stationary reductions of KdV can be solved 
by separation of variables is well-known (see, e.g., \cite{DKN}).
This classical method has recently found a lot of interesting new applications,
as shown in the survey \cite{Sk}.}. It can be shown that they are rather special 
equations: They 
are Hamiltonian with respect to a Poisson pair; this Poisson pair allows to
foliate the manifold $M_5$ into four--dimensional symplectic leaves with
special properties; each
symplectic leaf $S_4$ carries a Lagrangian foliation to which the vector
fields~\rref{eq:1.30} and~\rref{eq:1.31}  are tangent; the Poisson pair
defines a special set of coordinates on each $S_4$; in these coordinates the
Hamilton--Jacobi equations associated with the Hamiltonian
equations~\rref{eq:1.30} and~\rref{eq:1.31} can be simultaneously
solved by additive separation of variables. Most of these properties will be
proved in the next lecture.
\subsection{The modified KdV hierarchy}\label{lect:14}
We leave for the moment the theme of the reduction, and come back to the KdV
hierarchy in its general form. We notice that the first equations 
\rref{eq:1.20} can also be written in the form
\begin{equation}
  \label{eq:1.32}
  \begin{split}
\Bdpt{u}{1}&=(Q_1-\la Q_0)\al_0\\
\Bdpt{u}{3}&=(Q_1-\la Q_0)(\la\al_0+\al_1)\\
\Bdpt{u}{5}&=(Q_1-\la Q_0)(\la^2\al_0+\la\al_1+\al_2)\\
\end{split}
\end{equation}
This representation shows that these equations are Hamiltonian with respect to
the whole Poisson pencil. This elementary property can be exploited to
simply construct the modified KdV hierarchy. Let us write in general
\begin{equation}
  \label{eq:1.33}
  \Bdpt{u}{2j+1}=(Q_1+\la Q_0)\ala{j} (\la)\ ,
\end{equation}
where
\begin{equation}
  \label{eq:1.34}
  \ala{j}(\la)=\la^j\al_0+\la^{j-1}\al_{1}+\cdots+\al_j\ .
\end{equation}
By means of the Miura map~\rref{eq:1.6} we pull--back the one--forms
$\ala{j}$ to one--form $\bb{j}$ defined on
the phase space of the modified equation according to the
transformation law of one-forms,
\begin{equation}
  \label{eq:1.35}
  \beta(h)={\Phi^{\prime}_h}^* \al(\Phi(h))\ .
\end{equation}
We then define the equations
\begin{equation}
  \label{eq:1.36}
  \Bdpt{h}{2j+1}=P_h \bb{j}(\la)\ .
\end{equation}
They are $\Phi$--related to the corresponding equations of the KdV hierarchy,
exactly as the mKdV {\em equation} is $\Phi$--related to the KdV {\em
  equation}. Indeed,
\begin{equation}
  \label{eq:1.37}
  \Bdpt{u}{2j+1}=\Phi^\prime_h  \Bdpt{h}{2j+1}=\Phi^\prime_h
  {\Phi^{\prime}_h}^* \ala{j}(\Phi(h);\la)=Q_u\ala{j}(u;\la)\ . 
\end{equation}
It is therefore natural to call equations \rref{eq:1.36} 
the {\em  modified KdV hierarchy}. By using the explicit form of the operators 
$P_h$ and ${\Phi^{\prime}_h}^*$, it is easy to check that the modified
hierarchy is defined by the  conservation laws
\begin{equation}
  \label{eq:1.38}
  \Bdpt{h}{2j+1}=\del_x\Ha{2j+1},
\end{equation}
where 
\begin{equation}
  \label{eq:1.39}
 \Ha{2j+1}=-\frac12\ala{j}_x+\ala{j}h\>\ . 
\end{equation}
 \begin{exer} Write down explicitly the first three equations of the modified
  hierarchy.
\hfill $\square$\end{exer}
The above formulas are basic in the Sato approach. In the fourth lecture,
after a more accurate analysis of the Hamiltonian structure of the KdV
equations, we shall be led to consider the currents $\Ha{2j+1}$ as defining
a point of an infinite--dimensional Grassmannian. This point evolves in time
as the point $u$ moves according to the KdV equation. We shall determine the
equation of motion of the currents $\Ha{2j+1}$. They define a ``bigger''
hierarchy called the {\em Central System}. This system contains the KdV \ger y
as a particular reduction. It enjoys the property of being linearizable. In
this way, by a continuous process of extension motivated by the Hamiltonian
structure of the equations (from the single KdV equation to the KdV hierarchy 
and
to the Central System), we arrive at the result that the KdV flow can be
linearized. At this point the following picture of the possible strategies for 
solving the KdV equations emerges: either we pass to the
Sato infinite--dimensional Grassmannian and we use a linearization technique,
or we restrict the equation to a finite--dimensional invariant submanifold 
and we use a technique of separation of variables. The two strategies
complement themselves rather well. Our attitude is to see the Grassmannian
picture as a compact way of defining the equations, and the ``reductionist''
picture as an effective way for finding interesting classes of solutions.
\subsection*{The plan of the lectures} 
This is the web of ideas which we would like to make more precise in 
the following lectures. As cornerstone of our presentation we choose the
concept of Poisson pairs. In the second lecture, we develop the theory of these 
pairs up to the point of giving a sound basis to the concept of Lenard
chain. In the third lecture we exhibit a first class of examples, and we
explain a reduction technique allowing to construct the
Poisson pairs of the reduced flows. In the fourth lecture we give a second
look at the KdV theory, and we explain the reasons which, according to the
Hamiltonian standpoint, suggest to pass on the infinite--dimensional Sato
Grassmannian. In the fifth lecture we better explore the relation between the
two strategies, and we touch the concept of Lax representation. Finally, the
last lecture is devoted to the method of separation of variables. The
purpose is to show how the geometry of the reduced Poisson pairs can be 
used to  define the separation coordinates.
\newpage
\section{The method of Poisson pairs}\label{lect:2}
In the previous lecture we have outlined the birth of the method of Poisson
pairs and its main purpose: To define integrable \ger ies of vector fields.  
In this lecture we dwell on the theoretical basis of this construction
presenting the concept of Gel'fand--Zakharevich system.

The starting point is the notion of {\em Poisson manifold}. A manifold is said 
to be a Poisson manifold\footnote{The books \cite{LM} and \cite{Vaisman} are 
very good references for this topic.} if a composition law 
on scalar functions has been
defined obeying the usual properties of a Poisson bracket: bilinearity,
skewsymmetry, Jacobi identity and Leibnitz rule. The last condition 
means that that the Poisson bracket is  a derivation in each entry:
\begin{equation}
  \label{eq:2.1}
  \parpo{fg}{h}=\parpo{f}{h}g+f\parpo{g}{h}\ .
\end{equation}Therefore, by fixing the argument of one of the two entries and
keeping free the remaining one, we obtain a vector field,
\begin{equation}
  \label{eq:2.2}
  X_h=\parpo{\cdot}{h}\ .
\end{equation}
It is called the {\em Hamiltonian vector field} associated with the function
$h$ with respect to the given Poisson bracket.  Due to the remaining
conditions on the Poisson bracket, these vector fields are closed with respect 
to the commutator. They form a Lie algebra homeomorphic to the algebra of
functions defined by the Poisson bracket:
\begin{equation}
  \label{eq:2.3}
  [X_f,X_g]=X_{\parpo{f}{g}}\ .
  \end{equation}
Therefore a Poisson bracket on a manifold has a twofold role: it defines a Lie 
algebra structure on the ring of $C^\infty$--functions, and provides a 
representation of this algebra on the manifold by means of
the Hamiltonian vector fields.

Instead of working with the \parp, it is often suitable to work (especially in 
the
infinite--dimensional case) with the associated Poisson tensor. It is the
bivector field $P$ on $M$ defined by
\begin{equation}
  \label{eq:2.4}
  \parpo{f}{g}=\langle df, \, P\, dg\rangle\ .
\end{equation}
In local coordinates, its components $P^{jk}(x^1,\ldots,x^n)$ are the Poisson
brackets of the coordinate functions,
\begin{equation}
  \label{eq:2.5}
  P^{jk}(x^1,\ldots,x^n)=\parpo{x^j}{x^k}\ .
\end{equation}
By looking at this bivector field as a linear skewsymmetric map $P:T^*M\to TM$,
we can define the Hamiltonian vector fields $X_f$ as the images through $P$  
of the exact one-forms,
\begin{equation}
  \label{eq:2.6}
  X_f=P\,d f\ .
\end{equation}
In local coordinates this means
\begin{equation}
  \label{eq:2.7}
  X^j_f=P^{jk}\frac{\del f}{\del x^k}\ .
\end{equation}
\begin{exer}
Show that the components of the Poisson tensor satisfy the cyclic condition
\begin{equation}
  \label{eq:2.8}
  \sum_l \left(P^{jl}\frac{\del P^{km}}{\del x^l}+P^{kl}\frac{\del P^{mj}}{\del 
x^l}+
P^{ml}\frac{\del P^{jk}}{\del x^l}\right)=0\ .
\end{equation}
\hfill $\square$\end{exer}
\begin{exer}
Suppose that $M$ is an affine space $A$. Call $V$ the vector space associated
with $A$. Define a bivector field on $A$ as a mapping $P: A\times V^*\to V$
which satisfies the skewsymmetry condition 
\begin{equation}
  \label{eq:2.9}
  \langle \alpha,\, P_u\beta\rangle =- \langle \beta,\, P_u\al\rangle
\end{equation}
for every pair of covector $(\al,\beta)$ in $V^*$ and at each point $u\in\, A$.
Denote the directional derivative at the point $u$ of the mapping $u\mapsto 
P_u\al$ 
along the vector $v$ by
\begin{equation}
  \label{eq:2.9bis}
  P^\prime_u(\al;v)=\frac{d}{dt}P_{u+tv}\al\vert_{t=0}\ .
\end{equation}
Show that the bivector $P$ is a Poisson bivector if and only if it satisfies
the cyclic condition
\begin{equation}
  \label{eq:2.10}
  \langle \alpha,\, P^\prime_u(\beta;P_u\gamma)\rangle+\langle \beta,\,
  P^\prime_u(\gamma;P_u\alpha)\rangle+\langle \gamma,\,
  P^\prime_u(\alpha;P_u\beta)\rangle=0\ . 
\end{equation}
\hfill $\square$\end{exer}
\begin{exer}
Check that the bivector $Q_\la$ of equation \rref{eq:1.16}, associated with the
KdV equation, fulfills the conditions \rref{eq:2.9} and \rref{eq:2.10}.
\hfill $\square$\end{exer}
No condition is usually imposed on the rank of the Poisson bracket, that is, on 
the dimension of the vector space spanned by the Hamiltonian vector fields at 
each point of the
manifold. If these vector fields span the whole tangent space the bracket is
said to be regular, and the manifold $M$ turns out to be a symplectic
manifold. Indeed there exists, in this case, a unique symplectic
2-form $\omega$ such that
\begin{equation}
  \label{eq:2.11}
  \parpo{f}{g}=\omega(X_f,X_g)\ .
\end{equation}
More interesting is the case where the bracket is singular. In this case, the
Hamiltonian vector fields span a proper distribution $D$ on $M$. 
It is involutive but, generically, not of constant rank. 
Nonetheless, this distribution is {\em
completely integrable}: at each point there exists an integral submanifold of
maximal dimension which is tangent to the distribution. These submanifolds are 
symplectic manifolds, and are called the symplectic leaves of the Poisson
structure. The symplectic form is still defined by equation \rref{eq:2.11}. 
Indeed, 
even if there is not a 1--1 correspondence between (differentials of)
functions and Hamiltonian vector fields, this formula keeps its meaning, since 
the value of the Poisson bracket does not depend on the particular choice of
the Hamiltonian function associated with a given Hamiltonian vector field. We
arrive thus at the following conclusion: a Poisson manifold is either a
symplectic manifold or a {\em stratification\/} of symplectic manifolds
possibly of different dimensions. 
It can be proven  
that, in a sufficiently small open set where the rank of the Poisson tensor is 
constant, these 
symplectic manifolds are the level sets of 
some smooth functions $F_1,\ldots, F_k$, whose differentials span the kernel of
the Poisson tensor. They are called {\em Casimir functions\/} of $P$ (see 
below). 
\begin{exer}
Let $\{x_1,x_2,x_3\}$ be Cartesian coordinates in $M=\RR^3$. Prove that the
assignment 
\begin{equation}
  \label{eq:new}
  \parpo{x_1}{x_2}=x_3,\quad \parpo{x_1}{x_3}=-x_2,\quad\parpo{x_2}{x_3}=x_1
\end{equation}
defines a Poisson tensor on $M$. Find its Casimir function, and describe  
the symplectic foliation associated with it.
\hfill $\square$\end{exer}
After these brief preliminaries on Poisson manifolds as natural settings for the 
theory of Hamiltonian vector
fields, we pass to the theory of {\em bi-Hamiltonian\/} manifolds. Our purpose 
is
to provide evidence that they are a suitable setting for the theory of {\em
  integrable} Hamiltonian vector fields. The simplest connection between the
theory of integrable Hamiltonian vector fields and the theory of \varb\ is
given by the Gel'fand--Zakharevich (GZ) 
theorem~\cite{GZ93,GZ99} we shall discuss in this lecture.

A \varb\ is a Poisson manifold endowed with a {\em pair} of {\em compatible}
Poisson brackets. We shall denote these brackets with $\parpo{f}{g}_0$ and
$\parpo{f}{g}_1$. They are compatible if the Poisson pencil
\begin{equation}
  \label{eq:2.12}
  \parpo{f}{g}_\lambda:=\parpo{f}{g}_1-\la \parpo{f}{g}_0
\end{equation}
verifies the Jacobi identity for any value of the continuous (say, real) 
parameter $\la$. 
By means of this concept we catch the main features of the situation
first met in the KdV example of Lecture~\ref{lect:1}. The new feature
deserving attention is the dependence of the Poisson bracket~\rref{eq:2.12} on 
the parameter $\la$. It influences all the objects so far introduced on a
Poisson manifold: Hamiltonian fields and symplectic foliation. In particular,
this foliation changes with $\la$. The useful idea is to extract from this
moving foliation the invariant part. It is defined as the intersection of the
symplectic leaves of the pencil when $\la$ ranges over $\RR \cup
\{\infty\}$. The GZ theorem describes the structure of these intersections in
particular cases.

Let us suppose that the dimension of $M$ is odd, $\mbox{dim}\,M=2n+1$, and
that the rank of the Poisson pencil is maximal. This means that the dimension 
of the characteristic distribution spanned by the \ham\ \vefi\ is $2n$ for 
almost all the values of the parameter $\la$, and almost everywhere on the
manifold $M$. In this situation the generic symplectic leaf of the pencil has 
accordingly dimension $2n$ and the intersection of all the symplectic leaves
are submanifolds of dimension $n$. For brevity, we shall call this
intersection the {\em support\/} of the pencil. The GZ theorem displays 
an important property of the leaves of the support of the pencil.
\begin{theorem} On a $(2n+1)$--dimensional \varb, whose Poisson pencil has
  maximal rank, the leaves of the support are generically Lagrangian
  submanifolds of dimension $n$ contained on each symplectic leaf of dimension 
  $2n$.
\end{theorem}
This theorem contains two different statements. First of all it states that
the dimension of the support is exactly half of the dimension of the
generic symplectic leaf. It is the ``hard'' part of the theorem. Then it claims 
that the leaves of the support are Lagrangian submanifolds. Contrary to the
appearances, this is the easiest part of the theorem, as we shall see. To
better understand the content of the GZ theorem, we deem suitable to look at
it from a different and, so to say, more constructive, point of view. It
requires the use of the concept of {\em Casimir function\/},
defined as a function which commutes with all the other functions with respect
to the Poisson bracket. Equivalently, it can be defined as a function whose
differential belongs to the kernel of the Poisson tensor, i.e., a function
generating the null vector field. In the case of a Poisson pencil, the Casimir 
functions depend on the parameter $\la$. If the rank of the Poisson pencil is
maximal, the Casimir function is essentially unique (two Casimir functions are 
functionally dependent). The main content of the GZ theorem is that there
exists a Casimir function depending polynomially on the parameter $\la$, and
that the degree of this polynomial is exactly $n$ if $\mbox{dim}\,
M=2n+1$. Thus we can write the Casimir function in the form
\begin{equation}
  \label{eq:2.13}
  C(\la)=C_0\la^{n}+C_1\la^{n-1}+\cdots+C_n\ .
\end{equation}This result means that the Poisson pencil selects $n+1$
distinguished functions $(C_0,C_1,\ldots C_n)$. Generically these functions
are independent. Their common level surfaces are the leaves of the support of
the pencil. Indeed, on the support the function $C(\la)$ must be constant
independently of the particular value of $\la$. Thus all the coefficients
$(C_0,C_1,\ldots C_n)$ must be separately constant. Furthermore, as a
consequence of the fact that $C(\la)$ is a Casimir function, it is easily seen
that the coefficients $C_k$ verify the Lenard recursion relations,
\begin{equation}
  \label{eq:2.14}
  \parpo{\cdot}{C_k}_1=\parpo{\cdot}{C_{k+1}}_0\ ,
\end{equation}
together with the vanishing conditions
\begin{equation}
  \label{eq:2.15}
  \parpo{\cdot}{C_0}_0=\parpo{\cdot}{C_{n}}_1=0 \ .
\end{equation}
In the language of the previous lecture, the functions $(C_0,C_1,\ldots C_n)$
form a {\em Lenard chain}. A typical property of these functions is to be
mutually in involution:
\begin{equation}
  \label{eq:2.16}
\parpo{C_j}{C_k}_0=\parpo{C_j}{C_{k}}_1=0\ .  
\end{equation}
This is proved by repeatedly using the recursion
relation~\rref{eq:2.14} to go back and forth along the chain. It follows that 
the 
leaves of the support are isotropic submanifolds, but since they are of maximal 
dimension $n$ they are actually Lagrangian submanifolds. 
These short remarks should give a sufficiently detailed idea of the meaning of
the GZ theorem.
\begin{exer}
Check that that the integrals of motion $H_0$ and $H_1$ of the reduced flow
$X_1$ on the invariant submanifold $M_3$ considered in Section~\ref{lect:13}
are the coefficients of the Casimir function $C(\la)=\la H_0+H_1$ of the
Poisson pencil defined on $M_3$. 
\hfill $\square$\end{exer}
\begin{exer}
Prove the claim~\rref{eq:2.16} about the involutivity of the coefficients of 
a Casimir polynomial.
\hfill $\square$\end{exer}
From our standpoint, the above results are worthwhile of interest for two
different reasons: First of all they show how the Lenard recursion relations,
characteristic of the theory of ``soliton equations'', arise in a
theoretically sound way in the framework of \varb. Secondly, they highlight
the existence of a distinguished set of \ham\ $(C_0,C_1,\ldots C_n)$ on the
manifold $M$. Let us now choose one of the brackets of the pencil, say the
bracket $\parpu_0$. The function $C_0$ is a Casimir function for this bracket, 
and therefore its level surfaces are the synplectic leaves of the bracket
$\parpu_0$. Let us call $\omega_0$ the symplectic 2--form defined on these
submanifolds. As a consequence of the involution relation~\rref{eq:2.16}, the
restrictions of the $n$ functions  $(C_1,\ldots C_n)$ to the symplectic leaf
are in involution with respect to $\omega_0$. According to the
Arnol'd--Liouville theorem, they define a family (or ``hierarchy'') of $n$
completely integrable \ham\ \vefi s on the symplectic leaf.
\begin{defi}
The family of completely integrable \ham\ systems defined by the functions
$(C_1,\ldots C_n)$ on each symplectic leaf of the Poisson bracket $\parpu_0$
will be called the GZ \ger y associated with the Poisson pencil $\parpu_\la$
defined on the \varb\ $M$.
\end{defi}
We shall be particularly interested in the study of this \ger y for two
reasons. First we want to show that the previous simple concepts allow to
reconstruct a great deal of the KdV hierarchy, up to the linearization process 
on the infinite--dimensional Sato Grassmannian. In other words, we want to
show that the theory of Poisson pairs is a natural gate to the theory of
infinite-dimensional \ham\ systems described by partial differential equations 
of evolutionary type. Secondly, in a finite-dimensional setting, we want to show
that the GZ \vefi s are often more than integrable in the Liouville sense. 
Indeed, under some mild additional assumptions on the Poisson pencil, they are
separable, and the separation coordinates are dictated by the geometry of the
\varb. This result strenghtens the connection between Poisson pairs and
integrability.
\newpage 
\section{A first class of examples and the reduction technique}\label{lect:3}
The aim of this lecture is to present a first class of nontrivial examples of 
GZ \ger ies. The examples are constructed to reproduce the reduced KdV flows
discussed in the first lecture. The relation, however, will not be
immediately manifest, and the reader has to wait until the fifth lecture for
a full understanding of the motivations for some particular choice 
herewith made.

This lecture is split into three parts. In the first one we introduce a
simple class of \varb s called Lie--Poisson manifolds. They are duals of Lie
algebras endowed with a special Poisson pencil of Lie-theoretical origin. The
\ham\ \vefi s defined on these manifolds admit a {\em Lax representation} with
a Lax matrix depending linearly on the parameter $\la$. 
In the second part we show how to combine several copies 
of these manifolds, in 
such a way to obtain \ham\ \vefi s admitting a Lax representation depending
polynomially on the parameter $\la$. Finally, in the third part, we introduce 
the geometrical technique of reduction of Marsden and Ratiu. It will allow us to
specialize the form of the Lax matrix. The contact with the KdV theory, to be
done in the fifth lecture, will then consist in showing that the reduced KdV
flows admit exactly the Lax representation of the \ham\ \vefi s considered in
this lecture. This will ascertain the \bih\ character of the reduced
KdV flows. The lecture ends with an example worked out in detail.
\subsection{Lie--Poisson manifolds}\label{lect:31}
In this section $M=\algd$ is the dual of a Lie algebra $\alg$. We denote by
$S$ a point in $M$, and by $\ddd{F}{S}$ the differential of a function
$F:M\to\RR$. 
This differential is the unique element of the algebra $\alg$ such that
\begin{equation}
  \label{eq:3.1}
  \frac{dF}{dt}=\Big\langle \ddd{F}{S}, \dot{S}\Big\rangle
\end{equation}
along any curve passing through the point $S$. The Poisson pencil on $M$ is
defined by
\begin{equation}
  \label{eq:3.2}
  \parpo{F}{G}_\la=\Big\langle S+\la\, A, \left[\ddd{F}{S},\ddd{G}{S}\right] 
\Big\rangle\ ,
\end{equation}
where $A$ is any fixed element in $\algd$. In all the examples related to the
KdV theory, $\alg=\fraksl(2)$, $S$ and $\ddd{F}{S}$ are traceless $2\times 2$
matrices, and
\begin{equation}
  \label{eq:3.3}
  A=\mat2{0}{0}{1}{0}\ .
\end{equation}
The \ham\ \vefi\ $X_F$ has the form
\begin{equation}
  \label{eq:3.4}
  \dot{S}=\left[S+\la\,A, \ddd{F}{S}\right]\ .
\end{equation}
It is already in Lax form, with a Lax matrix given by
\begin{equation}
  \label{eq:3.5}
  L(\la)=\la\,A+S\ .
\end{equation}
\begin{exer} Compute the Poisson tensor and the \ham\ \vefi s associated with
  the pencil~\rref{eq:3.2}.
\hfill $\square$\end{exer}
\subsection{Polynomial extensions}\label{lect:32}
We consider two copies of the algebra $\alg$. Accordingly, we denote by
$(S_0,S_1)$ a point in $M$ and by $\left(\ddd{F}{S_0},\ddd{F}{S_1}\right)$ the 
differential of the function $F:M\to\RR$. By definition, along any curve 
$t\mapsto (S_0(t),S_1(t))$ we have
\begin{equation}
  \label{eq:3.6}
   \DT{F}=\Big\langle \ddd{F}{S_0}, \dot{S_0}\Big\rangle+ 
\Big\langle \ddd{F}{S_1}, \dot{S_1}\Big\rangle\ .
\end{equation}
The two copies of the algebra are intertwined by the Poisson brackets. As a
Poisson pair on $M$ we choose the following brackets
\begin{equation}
  \label{eq:3.78}
  \begin{split}
\parpo{F}{G}_0&=\Big\langle  A,
\left[\ddd{F}{S_0},\ddd{G}{S_1}\right]+\left[\ddd{F}{S_1},\ddd{G}{S_0}\right]
\Big\rangle+ \Big\langle S_1,
\left[\ddd{F}{S_0},\ddd{G}{S_0}\right]\Big\rangle\\
\parpo{F}{G}_1&=\Big\langle  A,
\left[\ddd{F}{S_1},\ddd{G}{S_1}\right]\Big\rangle- \Big\langle S_0,
\left[\ddd{F}{S_0},\ddd{G}{S_0}\right]\Big\rangle\end{split}
\end{equation}
The motivations can be found for instance in \cite{Magri-Magnano} (see also 
\cite{RSTS}). Later on we
shall see how to extend this definition to the case of an arbitrary number of
copies. 
\begin{exer} Check that equations~\rref{eq:3.78} indeed define a Poisson pair.
\hfill $\square$\end{exer}
Let us now study the \ham\ \vefi s. Those defined by the brackets $\parpu_0$
have the form
\begin{equation}
  \label{eq:3.9}
  \begin{split}
\dot{S_0}&=\left[A, \ddd{F}{S_1}\right]+\left[S_1, \ddd{F}{S_0}\right]\\
\dot{S_1}&=\left[A, \ddd{F}{S_0}\right]\ .
\end{split}
\end{equation}
Those defined by the second bracket $\parpu_1$ are
\begin{equation}
  \label{eq:3.10}
  \begin{split}
\dot{S_0}&=-\left[S_0, \ddd{F}{S_0}\right]\\
\dot{S_1}&=\left[A, \ddd{F}{S_1}\right]\ .
\end{split}
\end{equation}
It turns out that the \ham\ \vefi s associated with the Poisson pencil are
given by
\begin{equation}
  \label{eq:3.11}
    \begin{split}
\dot{S_0}&=-\left[S_0+\la\,S_1, \ddd{F}{S_0}\right]-\left[\la\,A, 
\ddd{F}{S_1}\right] \\
\dot{S_1}&=-\left[\la\, A, \ddd{F}{S_0}\right]+\left[A, \ddd{F}{S_1}\right] \ .
\end{split}
\end{equation}
This computation allows to display an interesting property of these \vefi
s. If we multiply the second equation by $\la$ and add the result to the
first equation we find
\begin{equation}
\label{eq:3.12}  
(\la^2\,A+\la\,S_1+S_0)^\bullet=\left[\ddd{F}{S_0},\la^2\,A+\la\,S_1+S_0
\right]
\ .
\end{equation}
This is a Lax representation with Lax matrix $L(\la)=\la^2\,A+\la\,S_1+S_0.$
It depends polynomially on the parameter of the pencil. We have thus
ascertained that all the \ham\ \vefi s relative to the Poisson
pencil~\rref{eq:3.11} admit a Lax representation. The converse, however, is
not necessarily true. Indeed, it must be noticed that the single Lax
equation~\rref{eq:3.12} is not sufficient to completely reconstruct the
Poisson pencil~\rref{eq:3.11}. Additional constraints on the matrix $L(\la)$
are required to make the problem well-posed. The kind of constraints to 
be set are suggested by the geometric theory of reduction which we shall now 
outline.
\subsection{Geometric reduction}\label{lect:33}
We herewith outline a specific variant~\cite{CMP} of the reduction technique of 
Marsden
and Ratiu~\cite{MR86} for Poisson manifolds. This variant is particularly
suitable for \varb s. 

Among the geometric objects defined by a Poisson pair $(P_0,P_1)$ on a
manifold $M$ we consider:
\begin{description}
\item[i)] a symplectic leaf $S$ of one of the two Poisson bivectors, say
  $P_0$.
\item[ii)] the annihilator $(TS)^0$ of the tangent bundle of $S$, spanned by
  the 1--forms vanishing on the tangent spaces to $S$.
\item[iii)] the image $D=P_1( TS)^0$ of this annihilator according to the
  second Poisson bivector $P_1$. It is spanned by the \ham\ \vefi s
  associated with the Casimir functions of $P_0$ by $P_1$.
\item[iv)] the intersection $E=D\cap TS$ of the distribution $D$ with the
  tangent bundle of the selected symplectic leaf $S$.
\end{description}
It can be show that $E$ is an integrable distribution as a consequence of the
compatibility of the Poisson brackets. Therefor we can consider the space of
leaves of this distribution, $N=S/E$. We assume $N$ to be a smooth
manifold. By the Marsden--Ratiu theorem, $N$ is a reduced \varb.

The reduced brackets on $N$ can be computed by using the process of
``prolongation of functions'' from $N$ to $M$. Given any function $f:N\to\RR$, 
we consider it as a function on $S$, invariant along the leaves of $E$. Then
we choose any function $F:M\to\RR$ which annihilates $D$ and coincides with
$f$ on $S$. This function is said to be a prolongation of $f$. It is not
unique, but this fact is not disturbing. It can be show that, if $F$ and $G$
are prolongations of $f$ and $g$, their bracket $\parpo{F}{G}_\la$ is an
invariant function along $E$. Therefore it defines a function on $N$ which is
by definition the reduced bracket $\parpo{f}{g}_\la$. The final result, of
course, is independent of the particular choices of the prolongations $F$ and
$G$.

\subsection{An explicit example}\label{lect:34}
According to the spirit of these lectures, rather than discussing the
proof of the reduction theorem stated in Section~\ref{lect:33}, we prefer to
illustrate it on a concrete example. Let us  thus perform the reduction of the 
Poisson pencil defined on two copies of $\alg=\fraksl(2)$. The matrices $S_0$
and $S_1$ are traceless matrices whose entries we denote as follows:
\begin{equation}
  \label{eq:3.13}
  S_0=\mat2{p_0}{r_0}{q_0}{-p_0}\ ,\qquad S_1=\mat2{p_1}{r_1}{q_1}{-p_1}\ .
\end{equation}
The space $M$ has dimension six, and the entries of $S_0$ and $S_1$ are global 
coordinates on it.  In these coordinates the differential of a function
$F:M\to\RR$ is represented by the pair of matrices
\begin{equation}
  \label{eq:3.14}
 \ddd{F}{S_0}=\mats2{\dsl{\frac12} \ddd{F}{p_0}}{\ddd{F}{q_0}}{\ddd{F}{r_0}}
{-\dsl{\frac12} \ddd{F}{p_0}}\ ,
\qquad  
\ddd{F}{S_1}=\mats2{\dsl{\frac12} \ddd{F}{p_1}}{\ddd{F}{q_1}}{\ddd{F}{r_1}}
{-\dsl{\frac12} \ddd{F}{p_1}}\ .
\end{equation}
\begin{exer} Define the pairing $\Big\langle \ddd{F}{S}, \dot{S}\Big\rangle$
  on $\alg$ as the trace of the product of the matrices $\ddd{F}{S}$ and
  $\dot{S}$. Show that the matrices~\rref{eq:3.14} verify the defining
  equation~\rref{eq:3.6}.
\hfill $\square$\end{exer}
The \ham\ \vefi s~\rref{eq:3.9} and~\rref{eq:3.10} are consequently given by
\begin{equation}
  \label{eq:3.15}
  \begin{split}
\dot{p_0}&=r_1\ddd{F}{r_0}-q_1\ddd{F}{q_0}-\ddd{F}{q_1}\\
\dot{q_0}&=q_1\ddd{F}{p_0}-2p_1\ddd{F}{r_0}+\ddd{F}{p_1}\\
\dot{r_0}&=2p_1\ddd{F}{q_0}-r_1\ddd{F}{p_0}\\
\dot{p_1}&=-\ddd{F}{q_0}\\
\dot{q_1}&=\ddd{F}{p_0}\\
\dot{r_1}&=0
\end{split}
\end{equation}
and by
\begin{equation}
  \label{eq:3.16}
  \begin{split}
\dot{p_0}&=-r_0\ddd{F}{r_0}+q_0\ddd{F}{q_0}\\
\dot{q_0}&=2p_0\ddd{F}{r_0}-q_0\ddd{F}{p_0}\\
\dot{r_0}&=-2p_0\ddd{F}{q_0}+r_0\ddd{F}{p_0}\\
\dot{p_1}&=-\ddd{F}{q_1}\\
\dot{q_1}&=\ddd{F}{p_1}\\
\dot{r_1}&=0
\end{split}
\end{equation}
respectively.
\vspace{3truemm}\\
{\bf Step 1: The reduced space $N$}.
\vspace{2truemm}\\ 
First we notice that the \ham\
\vefi s~\rref{eq:3.15} verify the constraints
\begin{equation}
  \label{eq:3.17}
  \dot{r_1}=0\ ,\qquad (r_0+p_1^2+r_1q_1)^\bullet=0\ .
\end{equation}
It follows that the submanifold $S\subset M$ defined by the equations
\begin{equation}
  \label{eq:3.18}
  r_1=1\ ,\qquad  r_0+p_1^2+r_1q_1=0
\end{equation}
is a symplectic leaf of the first Poisson bivector $P_0$.
Furthermore, it 
follows that the annihilator $(TS)^0$ is spanned by the exact 1--forms $dr_1$
and $d(r_0+p_1^2+r_1q_1)$. By computing the images of these forms according to 
the second Poisson bivector~\rref{eq:3.16}, we find the distribution $D$. It is 
spanned by the single vector field
\begin{equation}
  \label{eq:3.19}
  \begin{split}
\dot{p_0}&=-r_0\\
\dot{q_0}&=2p_0\\
\dot{r_0}&=0\\
\dot{p_1}&=-1\\
\dot{q_1}&=2p_1\\
\dot{r_1}&=0
\end{split}
\end{equation}
which verifies the five constraints
\begin{equation}
  \label{eq:3.20}
\begin{array}{l}
  (p_0-r_0p_1)^\bullet=0,\qquad (q_0+2p_0p_1-r_0p_1^2)^\bullet=0,\\
(q_1+p_1^2)^\bullet=0,\quad\dot{r_0}=0,\quad\dot{r_1}=0.\end{array}
\end{equation}
They show that $D\subset TS$, and therefore $E=D$.  Moreover they yield that the
leaves of $E$ on $S$ are the level curves of the functions
\begin{equation}
  \label{eq:3.21}
  \begin{split}
u_1&=q_1+p_1^2\\
u_2&=p_0+p_1q_1+p_1^3\\
u_3&=q_0+2p_0p_1+q_1p_1^2+p_1^4\\
\end{split}
\end{equation}
We conclude that:
\begin{itemize}
\item $N=\RR^3$;
\item $(u_1,u_2,u_3)$ are global coordinates on $N$;
\item the canonical projection $\pi:S\to S/E$ is defined by 
equations~\rref{eq:3.21}. 
\end{itemize}
{\bf Step 2: The reduced brackets.} 
\vspace{2truemm}\\
Consider any function $f:N\to\RR$. The function 
\begin{equation}
  \label{eq:3.22}
  F:=f(q_1+p_1^2,p_0+p_1q_1+p_1^3,q_0+2p_0p_1+q_1p_1^2+p_1^4)
\end{equation}
is clearly a prolongation of $f$ to $M$, since it coincides with $f$ on $S$,
and is invariant along $D$. We can thus use $F$ to compute the first component 
of the reduced \ham\ \vefi\ on $N$ according to the following algorithm:
\begin{equation}
  \label{eq:3.23}
  \begin{split}
\dot{u_1}&\eqcon{\ref{eq:3.21}}\dot{q_1}+2p_1\dot{p_1}\eqcon{\ref{eq:3.15}}\ddd{
F}{p_0}
-p_1\ddd{F}{q_0}\\ 
&\eqcon{\ref{eq:3.22}}\left(\ddd{f}{u_2}+2p_1\ddd{f}{u_3}\right)-2p_1
\ddd{f}{u_3}=\ddd{f}{u_2}\ .\\
\end{split}
\end{equation}
The other components are evaluated in the same way. The final result is that
the \ham\ \vefi s associated with the reduced Poisson pencil on $N$ are
defined by
\begin{equation}
  \label{eq:3.24}
  \begin{split}
&\dot{u_1}=(u_1+\la)\ddd{f}{u_2}+2u_2\ddd{f}{u_3}\\
&\dot{u_2}=-(u_1+\la)\ddd{f}{u_1}+(u_3-2\la u_1)\ddd{f}{u_3}\\
&\dot{u_3}=-2u_2\ddd{f}{u_1}+(2\la u_1-u_3)\ddd{f}{u_2}\\
\end{split}
\end{equation}
With this reduction process we passed from a six--dimensional manifold $M$ to a 
three dimensional manifold $N$. Later on, we shall see that this manifold
coincides with the invariant submanifold $M_3$ of KdV, defined by the
constraint
\begin{equation}
  \label{eq:3.25}
  u_{xxx}-6uu_x=0\ .
\end{equation}
{\bf Step 3: the GZ hierarchy}.
\vspace{2truemm}\\
To compute the Casimir function of the
pencil~\rref{eq:3.24} we notice that these vector fields obey the constraint
\begin{equation}
  \label{eq:3.26}
  (2\la u_1-u_3)\dot{u_1}+2u_2\dot{u_2}-(u_1+\la)\dot{u_3}=0\ .
\end{equation}
Therefore, integrating this equation, we obtain that
\begin{equation}
  \label{eq:3.27}
  C(\la)=\la(u_1^2-u_3)+(u_2^2-u_1u_3)=\la C_0+C_1
\end{equation}
is the Casimir sought for. It fulfills the scheme of the GZ theorem, and it
defines a ``short'' Lenard chain
\begin{equation}
  \label{eq:3.28}
  P_0 d C_0=0\qquad P_1 d C_0=P_0 d C_1=X_1\qquad P_1 d C_1=0\ .
\end{equation}
Therefore the GZ ``\ger y'' consists of the single vector field
\begin{equation}
  \label{eq:3.29}
  X_1\>:\qquad
\begin{array}{l} \dot{u_1}=2u_2\\
  \dot{u_2}=u_3+2u_1^2\\
  \dot{u_3}=4u_1u_2\end{array}
\end{equation}
As a last remark, we notice that this \vefi\ coincides with the restriction of 
the first equation $\Bdpt{u}{1}=u_x$ of the KdV \ger y on the invariant
submanifold~\rref{eq:3.25}. Indeed, by the procedure explained in
Section~\ref{lect:1}, the reduced equation written in the ``Cauchy data
coordinates'' $(u,u_x,u_{xx})$ is given by
\begin{equation}
  \label{eq:3.30}
  \begin{split}&\Bdpt{u}{1}=u_x\\
   & \Bdpt{u_x}{1}=u_{xx}\\
   & \Bdpt{u_{xx}}{1}=6uu_{x}
\end{split}
\end{equation}
We can now pass from~\rref{eq:3.30} to~\rref{eq:3.29} 
by the change of variables
\begin{equation}
  \label{eq:3.30b}
  u_1=\dsl\frac12 u,\quad u_2=\frac14 u_x,\quad u_3=\frac14u_{xx}-\frac12 u^2\ .
\end{equation}
This remark shows that the simplest reduced KdV flow is \bih. In the fifth
lecture we shall see that this property is general, and we shall explain the
origin of the seemingly ``ad hoc'' change of variables~\rref{eq:3.30b}. 
\subsection{A more general example}\label{lect:35}
To deal with higher--order reduced KdV flows, we have to extend the class of
\varb s to be considered. We outline the case of three copies of the algebra
$\alg$. The formulas are similar to the ones of equation \rref{eq:3.78}, 
albeit a little more involved. 
The brackets $\parpo{F}{G}_0$ and $\parpo{F}{G}_1$ are now given by
\begin{equation}
  \label{eq:3.31}
  \begin{split}
\parpo{F}{G}_0&=\Big\langle  A,
\left[\ddd{F}{S_0},\ddd{G}{S_2}\right]+\left[\ddd{F}{S_1},\ddd{G}{S_1}\right]+
\left[\ddd{F}{S_2},\ddd{G}{S_0}\right]\Big\rangle \\
&+\Big\langle S_2,
\left[\ddd{F}{S_0},\ddd{G}{S_1}\right]+\left[\ddd{F}{S_1},\ddd{G}{S_0}\right]
\Big\rangle\\
&+\Big\langle S_1,
\left[\ddd{F}{S_0},\ddd{G}{S_0}\right]\Big\rangle\\
\end{split}
\end{equation}
and
\begin{equation}
  \label{eq:3.32}
  \begin{split}
\parpo{F}{G}_1&=\Big\langle  A,
\left[\ddd{F}{S_1},\ddd{G}{S_2}\right]+\left[\ddd{F}{S_2},\ddd{G}{S_1}\right]
\Big\rangle \\
&+\Big\langle S_2,
\left[\ddd{F}{S_1},\ddd{G}{S_1}\right]
\Big\rangle\\
&-\Big\langle S_0,
\left[\ddd{F}{S_0},\ddd{G}{S_0}\right]\Big\rangle.\\
\end{split}
\end{equation}
The comparison of the two examples allows to infer by induction the general
rule for the Poisson pair, holding in the case of an arbitrary (finite) number 
of copies of $\alg$. 
The pencil \rref{eq:3.31}--\rref{eq:3.32} can be reduced according to the 
procedure shown
before. If $\alg=\fraksl(2)$ and $A$ is still given by~\rref{eq:3.3},
the final
result of the process is the following:  We start from a nine--dimensional
manifold $M$ and, after reduction, we arrive at a five--dimensional manifold
$N$. It fulfills the assumption of the GZ theorem. The GZ \ger y consists of
two \vefi s, which are the reduced KdV flows given by \rref{eq:1.30} and 
\rref{eq:1.31}.
\begin{exer}
Perform the reduction of the pencil \rref{eq:3.31}--\rref{eq:3.32}) for 
$\alg=\fraksl(2)$.\\
\null\hfill $\square$\end{exer}
\newpage
\section{The KdV theory revisited}\label{lect:4}
In this lecture we consider again the KdV theory, but from a new point of
view.  Our purpose is twofold.  The first aim is to show that the KdV \ger y
is another example of GZ \ger y. The second aim is to explain in which sense
the KdV \ger y can be linearized. The algebraic linearization procedure dealt
with in this lecture was suggested for the first time by Sato~\cite{SS82} (see 
also the developments contained in~\cite{DJKM,SW85,Ta89}), who exploited 
the so--called Lax representation of the KdV \ger y in the algebra of
pseudo--differential  operators. Here we shall give a different description, 
strictly
related to the \ham\ representation of the KdV \ger y as a kind of 
infinite-dimensional  GZ \ger y.  However, the presentation does not go beyond
the limits of a simple sketch of the theory. We refer to~\cite{fmp98} for full 
details.
\subsection{Poisson pairs on a loop algebra}\label{lect:41}
In this section we consider the infinite-dimensional Lie algebra $M$ of
$C^\infty$--maps from the circle $S^1$ into $\alg=\fraksl(2)$. A generic
point of this manifold is presently a $2\times 2$ traceless matrix
\begin{equation}
  \label{eq:4.1}
  S=\mat2{p(x)}{r(x)}{q(x)}{-p(x)}\ ,
\end{equation}
whose entries are periodic functions of the coordinate $x$ running over the
circle.  The three functions $(p,q,r)$ play the role of
``coordinates'' on our manifold. The scalar-valued functions $F:M\to\RR$ to
be considered are local functionals
\begin{equation}
  \label{eq:4.2}
  F=\int_{S^1} f(p,q,r;p_x,q_x,r_x;\dots)\, dx \ .
\end{equation}
As before, their differentials are given by the matrices
\begin{equation}
  \label{eq:4.3}
  \fddd{F}{S}=\mats2{\dsl{\frac12} \fddd{f}{p}}{\fddd{f}{q}}{\fddd{f}{r}}
{-\dsl{\frac12} \fddd{f}{p}}\ ,
\end{equation}
whose entries are the variational derivatives of the Lagrangian density $f$
with respect to  the functions $(p,q,r)$. The Poisson pencil is similar to the 
first one considered in the previous lecture (see equation \rref{eq:3.2}). It is
defined by
\begin{equation}
  \label{eq:4.4}
  \parpo{F}{G}_\la=\Big\langle S+\la\, A, \left[\fddd{F}{S},\fddd{G}{S}\right]
  \Big\rangle +\omega\left(\fddd{F}{S},\fddd{G}{S}\right).
\end{equation}
 It differs from the previous example by the addition of the nontrivial
 cocycle
 \begin{equation}
   \label{eq:4.5}
   \omega\left(a,b\right)=\int_{S^1}\frac{da}{dx}\, b\, dx \ .
 \end{equation}
This term is essential to generate partial differential equations.
It is responsible for the appearance of the
partial derivative in the expansion of the \ham\ \vefi s
  \begin{equation}
    \label{eq:4.6}
    \dot{S}=\left(\fddd{F}{S}\right)_x+\left[S+\la\,A, \fddd{F}{S}\right]\ .
  \end{equation}
\begin{exer}
Recall that a two--cocycle on $\alg$ is a bilinear skewsymmetric map
$\omega:\alg\times \alg\to\RR$ which verifies the cyclic condition
\begin{displaymath}
  \omega(a,[b,c])+ \omega(b,[c,a])+ \omega(c,[a,b])=0\ .
\end{displaymath}
Using this identity and the periodic boundary conditions check that
equation \rref{eq:4.6} defines a Poisson bivector.
\hfill $\square$\end{exer}
\subsection{Poisson reduction}\label{lect:42}
We apply the same reduction technique used in the previous lecture, avoiding
to give all the details of the computations. They can be either worked out by
exercise or found in~\cite{CMP, Pondi}

The first Poisson bivector $P_0$ is defined by
\begin{equation}
  \label{eq:4.7}
   \dot{S}=\left[A, \fddd{F}{S}\right] ,
\end{equation}
where $A$ is still defined by equation \rref{eq:3.3}. These \ham\ \vefi s obey 
the
only constraint
$\dot{r}=0$. Therefore the submanifold $\CS$ formed
by the matrices
\begin{equation}
  \label{eq:4.8}
    S=\mat2{p}{1}{q}{-p}
\end{equation}
is a symplectic leaf of $P_0$. The annihilator $(T\CS)^0$ is spanned by the
differentials of the
functionals $F:M\to\RR$ depending only on the coordinate function
$r$. Consequently, the distribution $D$ is spanned by the \vefi s
\begin{equation}
  \label{eq:4.9}
  \begin{split}
    &\dot{p}=\fddd{f}{r}\\
    &\dot{q}=\left(\fddd{f}{r}\right)_x-2p\fddd{f}{r}\\
    &\dot{r}=0\\
\end{split}
\end{equation}
The distribution $D$ is thus tangent to $\CS$ and $E$ coincides with $D$. The
\vefi~\rref{eq:4.9} verifies the constraint
\begin{equation}
  \label{eq:4.10}
  \dot{q}+2p\dot{p}+\dot{p_x}=(q+p^2+p_x)^\bullet=0\ .
\end{equation}
It follows that the leaves of the distribution $E$ are the level sets of the
function
\begin{equation}
  \label{eq:4.11}
  u=q+p^2+p_x \ .
\end{equation}
Therefore the quotient space $N$ is the space of scalar functions $u:S^1\to
\RR$, and ~\rref{eq:4.11} is the canonical projection $\pi:\CS\to\CS/E$. We see
that the manifold $N$ is (isomorphic to) the phase space of the KdV
equation.

We use the projection~\rref{eq:4.11} to compute the reduced Poisson
bivectors. The scheme of the computation is always the same. First we prolong
any functional $\CF=\int_{S^1}f(u,u_x,\cdots)dx$ on $N$ into the functional
\begin{equation}
  \label{eq:4.12}
  F(p,q,r)=\int_{S^1}f(q+p^2+p_x ,q_x+2pp_x+p_{xx};\dots)\, dx 
\end{equation}
on $\CS$. Then we compute its differential at the points of $\CS$,
\begin{equation}
  \label{eq:4.13}
   \fddd{F}{S}=\mats2{\dsl{-\frac12} 
\left(\fddd{F}{u}\right)_x+p\fddd{F}{u}}{\fddd{F}{u}}
{0}{\dsl{\frac12} \left(\fddd{F}{u}\right)_x-p\fddd{F}{u}} \ .
\end{equation}
Finally, we evaluate the reduced \ham\ \vefi s on $N$ according to the usual
scheme:
\begin{equation}
  \label{eq:4.14}
  \begin{split}
\dot{u}&\eqcon{4.11}\dot{q}+\dot{p_x}+2p\dot{p}\\
&\eqcon{4.6}\left[\left(\fddd{f}{u}\right)_x+(q+\la)\fddd{f}{p}-
2p\fddd{f}{r}\right]+
\left[\frac12\left(\fddd{f}{p}\right)_x+\fddd{f}{r}+(q+\la)\fddd{f}{q}
\right]_x\\
&\quad+2p
\left[\frac12\left(\fddd{f}{p}\right)_x+\fddd{f}{r}+(q+\la)\fddd{f}{q}
\right]\\
&\eqcon{4.13} (q+\la)\left[-\left(\fddd{f}{u}\right)_x+2p\fddd{f}{u}\right]+
\left[-
\frac12\left(\fddd{f}{u}\right)_{xx}+\left(p\fddd{f}{u}\right)_x+(q+\la)\fddd{f}
{u}\right]_x\\
&\quad
2p\left[-
\frac12\left(\fddd{f}{u}\right)_{xx}+\left(p\fddd{f}{u}\right)_x+(q+\la)
\fddd{f}{u}\right]\\
&=-
\frac12\left(\fddd{f}{u}\right)_{xxx}+2(q+p_x+p^2+\la)\left(\fddd{f}{u}
\right)_{x}+
(q_x+p_{xx}+2pp_x)\fddd{f}{u}\\
&\eqcon{4.11}-
\frac12\left(\fddd{f}{u}\right)_{xxx}+2(u+\la)\left(\fddd{f}{u}\right)_{x}+
u_x\fddd{f}{u}.
\end{split}
\end{equation}
We obtain the Poisson pencil of the KdV equation. This pencil is therefore the 
reduction of the ``canonical'' pencil~\rref{eq:4.4} over a loop algebra.
\subsection{The GZ  \ger y}\label{lect:43}
The simplest way for computing the Casimir function of the above pencil is to
use the Miura map. Since this map relates the pencil to the simple bivector of 
the mKdV equation, it is sufficient to compute the Casimir of the latter
bivector, and to transform it back to the phase space of the KdV equation.

We notice that the Casimir function of the mKdV \ger y \rref{eq:1.38} is given 
by
\begin{equation}
  \label{eq:4.15}
  \Hh(h)=2\,z\,\int_{S^1} h\, dx \ ,
\end{equation}
where the constant $z$ has been inserted for future convenience.

To obtain the Casimir function of the KdV equation, we must ``invert'' the Miura
map by expressing $h$ as a function of $u$. To do that we 
exploit the dependence of the Miura map on the
parameter $\la=z^2$ of the pencil. We know that in the finite-dimensional
case the Casimir function can be found as a polynomial in $\la$. In the
infinite-dimensional case, we expect the Casimir function to be represented
by a series. It is then natural to look at $h$ in the form of a Laurent
series in $z$, 
\begin{equation}
  \label{eq:4.16}
  h(z)=z+\sum_{l\ge 1}{h_l}{z^{-l}}\ ,
\end{equation}
whose coefficients $h_l$ are scalar-valued periodic functions of $x$. In this
way we change our point of view on the Miura map. Henceforth it must be looked 
at as a relation between a scalar function $u$ and a Laurent series
$h(z)$. This  change of perspective deeply influences all the mKdV theory. It 
is a possible starting point for the Sato picture of the KdV theory, as we
shall show later.

By inserting the expansion \rref{eq:4.16} into the Miura map $h_x+h^2=u+z^2$
and equating the coefficients of different powers of $z$, we easily compute
recursively the coefficients $h_l$ as differential polynomial of the function
$u$. The first ones are
\begin{equation}
\label{eq:4.17}
\begin{array}{l}
h_1=\frac12u\\
h_2=-\frac14u_x\\
h_3=\frac18(u_{xx}-u^2)\\
h_4=-\frac1{16}(u_{xxx}-4uu_x)\\
h_5=\frac1{32}(u_{xxxx}-6uu_{xx}-5u_x^2+2u^3)\ .
\end{array}
\end{equation}
One can notice (see~\cite{Alber}) that all the even coefficients
$h_{2l}$ are total $x$--derivatives. 
This remark explains the ``strange'' enumeration with odd times
used for the KdV \ger y in the first lecture.

To compute concretely the GZ vector fields, besides the Casimir function
\begin{equation}
  \label{eq:4.18}
  \Hh(u,z)=2z\sum_{l\ge 1}\int_{S^1} {h_l}{z^{-l}}\, dx\ ,
\end{equation}
we need its differential. To simplify the notation we set
\begin{equation}
  \label{eq:4.19}
 \al:= \fddd{\Hh}{u}=1+\sum_{l\ge 1} {\al_l}{z^{-l}}   \ .
\end{equation}
Once again, the simplest way for evaluating this series is to use the Miura
map. We notice that $\beta=2z$ is the differential of the Casimir of the mKdV
equation. From the transformation law of 1-forms,
\begin{equation}
  \label{eq:4.20}
  {\Phi^\prime_h}^*(\al)=\beta\ ,
\end{equation}
we then conclude that that $\al$ solves the equation
\begin{equation}
  \label{eq:4.21}
  -\al_x+2\al h=2z  \ .
\end{equation}
As before, the coefficients $\al_l$ can be computed recursively. One finds a
Laurent series in $\la=z^2$,
\begin{equation}
  \label{eq:4.22}
  \al=1-\frac12 u\la^{-1}+\frac18(3u^2-u_{xx})\la^{-2}+\cdots\ ,
\end{equation}
whose first coefficients have already appeared in \rref{eq:1.19}.
From $\al$ we can easily evaluate the Lenard partial sums
$\ala{j}=\big(\la^j\al\big)_+$ and write the odd GZ equations in the form
\begin{equation}
  \label{eq:4.23}
  \Bdpt{u}{2j+1}=\Big(-
\frac12\del_{xxx}+2(u+\la)\del_x+u_x\Big)\big(\ala{j}\big)  \ .
\end{equation}
The even ones are 
\begin{equation}
  \label{eq:4.24}
   \Bdpt{u}{2j}=0\ .
\end{equation}
The above equations completely and tersely define the KdV \ger y from the
standpoint of the method of Poisson pairs.
\subsection{The Central System}\label{lect:44}
We shall now pursue a little further the far--reaching consequences of the
change of point of view introduced in the previous
subsection. According to this new point of view, the mKdV \ger y is
defined in the space $\CL$ of the Laurent series in $z$ truncated form above. 
This
affects the whole picture.

Let us consider again the basic formulas of the mKdV theory. They are the
Miura map,
\begin{equation}
  \label{eq:4.25}
  h_x+h^2=u+z^2\> ,
\end{equation}
the formula for the currents~\rref{eq:1.39},
\begin{equation}
\Ha{2j+1}=-\frac12\ala{j}_x+\ala{j}h,\quad \Ha{2j}=0\>,  
\label{eq:4.26}
  \end{equation}
and the definition of the mKdV \ger y
\begin{equation}
  \label{eq:4.27}
  \Bdpt{h}{j}=\del_x\Ha{j}\ .
\end{equation}
They were obtained in the first lecture. Presently they are complemented by
the information that $h(z)$ is a Laurent series of the form~\rref{eq:4.16}. We 
shall now investigate the meaning of the above formulas in this new
setting. 

We start form the series $h(z)$, and we associate with it a new family of
Laurent series $\ha{j}(z)$ defined recursively by
\begin{equation}
  \label{eq:4.28}
  \ha{j+1}=\ha{j}_x+h\ha{j}\ ,
\end{equation}
starting from $\ha{0}=1$. They form a {\em moving frame} associated with the
point $h$ in the space of (truncated) Laurent series. The first three elements 
of this frame are explicitly given by
\begin{equation}
  \label{eq:4.29}
  \ha{0}=1,\qquad \ha{1}=h,\qquad \ha{2}=h_x+h^2\>\ .
\end{equation}
We see the basic block $h_x+h^2$ of the Miura transformation appearing. We call
$\CH_+$ the linear span of the series $\{\ha{j}\}_{j\ge 0}$. It is a linear 
subspace of $\CL$, attached to the point $h$. We can now interpret the three
basic formulas of the mKdV theory as properties of this linear space:
\begin{itemize}
\item The Miura map~\rref{eq:4.25} tells us that the linear space $\CH_+$ is
  invariant with respect to the multiplication by $\la$,
  \begin{equation}
    \label{eq:4.30}
    \la(\CH_+)\subset \CH_+\ .
  \end{equation}
\item The formula~\rref{eq:4.26} for the currents then entails that the
  currents $\Ha{j}$, for $j\in\,\NN$, belong to $\CH_+$:
  \begin{equation}
    \label{eq:4.31}
    \Ha{j}\in\,\CH_+\>\ .
  \end{equation}
\item Furthermore, in conjunction with equation \rref{eq:4.21}, it entails that 
the 
  asymptotic expansion of the currents $\Ha{j}$ has the form
  \begin{equation}
    \label{eq:4.32}
    \Ha{j}=z^j+\sum_{l\ge 1} {H^i_l}{z^{-l}}=z^j+O(z^{-1})\ .
  \end{equation}
\item Finally, the mKdV equations~\rref{eq:4.27} can be seen as the
  commutativity conditions of the operators $(\del_x+h)$ and
  $\big(\Bdpt{}{j}+\Ha{j}\big)$:
  \begin{equation}
    \label{eq:4.33}
    \left[\del_x+h,\Bdpt{}{j}+\Ha{j}\right]=0\>\ .
  \end{equation}
\end{itemize}
Used together,  conditions~\rref{eq:4.31} and~\rref{eq:4.33} imply that the 
operators
$\big(\Bdpt{}{j}+\Ha{j}\big)$ leave the linear space $\CH_+$ invariant:
\begin{equation}
  \label{eq:4.34}
  \big(\Bdpt{}{j}+\Ha{j}\big)(\CH_+)\subset \CH_+\>\ .
\end{equation}
This is the abstract but simple form of the laws governing the time evolution
of the currents $\Ha{j}$. These equations are the ``top'' of the KdV theory,
and form the basis of the Sato theory. It is not difficult to give them a
concrete form. By using the form of the expansion~\rref{eq:4.32} it is easy to 
show that equations \rref{eq:4.34} are equivalent to the infinite system of
Riccati--type equations on the currents $\Ha{j}$:
\begin{equation}
  \label{eq:4.35}
  \Bdpt{\Ha{j}}{k}+\Ha{j}\Ha{k}=\Ha{j+k}+\sum_{l=1}^j
  H^k_l\Ha{j-l}+\sum_{l=1}^k H^j_l\Ha{k-l}\ .
\end{equation}
It will be called the Central System (CS).
\begin{exer}
Prove formulas~\rref{eq:4.31} and~\rref{eq:4.32}.
\end{exer}
\subsection{The linearization process}\label{lect:45}
The first reward of the previous work is the discovery of a linearization
process.  The equations~\rref{eq:4.35} of the Central System are not directly
linearizable, but they can be easily transformed into a new system of
linearizable Riccati equations by a transformation in the space of
currents. This idea is realized once again by a ``Miura map''. The novelty,
however, is that this map is now operating on the space of currents rather than
on the phase space of the KdV equation.

We simply give the final result. Let us consider a new family of currents 
$\{\Wa{k}\}_{k\ge0}$
of the form
\begin{equation}
  \label{eq:4.36}
  \Wa{k}=z^k+\sum_{l\ge 1} {W^k_l}{z^{-l}}\>,
\end{equation}
and let us denote by $\CW_+$ their linear span in $\CL$. We define (see
also~\cite{Ta89}) a new system of equations on the currents $\Wa{k}$ by
imposing the ``constraints''
\begin{equation} \label{eq:4.37}
  \big(\Bdpt{}{k}+z^{k}\big)(\CW_+)\subset \CW_+
\end{equation}
on their linear span $\CW_+$.  It is easily seen that they take the explicit
form
\begin{equation}
  \label{eq:4.38}
  \Bdpt{\Wa{k}}{j}+z^j\Wa{k}=\Wa{j+k}+\sum_{l=1}^j W^k_l\Wa{j-l}\>\ .
\end{equation}
They will be called the {\em Sato equations} (on the ``big cell of the Sato
Grassmannian'').  They are a system of linearizable Riccati equations. This can 
be seen either from the geometry of a suitable group action on the
Grassmannian~\cite{DJKM} or by means of the following more elementary
considerations. We write equations \rref{eq:4.38} in the matrix form
\begin{equation}  \label{eq:4.39}
  \Bdpt{\Ww} {j}+ \Ww \cdot{}^T \La^k-\La^k\cdot\Ww=\Ww\Ga_k\Ww\ ,
\end{equation}
where $\Ww=\big(W^k_l\big)$ is the matrix of the components of the currents 
$\Wa{k}$,
$\La$ is the infinite shift matrix
\begin{equation}\label{eq:4.40}
\La=\left[\begin{array}{ccccc}
0 & 1 & 0 & \cdots  & \\
0 & 0  &  1 & 0 &\cdots \\
\vdots & & \ddots & \ddots &\\
\vdots & & & \ddots & \ddots \\
\vdots & & &  & \ddots
\end{array}\right]\ ,
\end{equation}
and $\Ga^k$ is the convolution matrix of level $k$,
\begin{equation}\label{eq:4.41}
\Gamma_k=\left[\begin{array}{cccccc}
0 &  \cdots & & 1 & 0 & \cdots\\
\vdots & & 1 & 0 & \cdots & \cdots\\
& \cdot & \cdot & & & \\
1 & 0 & & & & \\
\vdots & & & & &
\end{array}\right]\ .
\end{equation}
One can thus check that the matrix Riccati equation~\rref{eq:4.39} is solved
by the matrix
\begin{equation}
  \label{eq:4.42}
  \Ww=\Vv\cdot \Uu^{-1}\ ,
\end{equation}
where $\Uu$ and $\Vv$ satisfy the
constant coefficients linear system
\begin{equation}\label{eq:4.43}
\Bdpt{}{k} \Uu={}^T\La^k \Uu-\Ga_k\Vv\ , \qquad \Bdpt{}{k} \Vv=\La^k\Vv\ .
\end{equation} 
The closing remark is that the Sato equations are mapped into the Central
System~\rref{eq:4.35} by the following algebraic Miura map:
\begin{equation}
  \label{eq:4.44}
  \Ha{j}=\frac{\sum_{l=0}^j W^0_{j-l}\Wa{l}}{\Wa{0}}\ .
\end{equation}
The outcome of this long chain of extensions and transformations is the
following algorithm for solving the KdV equation:
\begin{description}
\item[i)] First we solve the linear system~\rref{eq:4.43}, with a suitably
  chosen initial condition, which we do not discuss here;
\item[ii)] Then we use the projective transformation~\rref{eq:4.42} and the
  Miura map~\rref{eq:4.44} to recover the currents $\Ha{j}$;
\item[iii)] Finally, we extract the first current $\Ha{1}=h$, and we evaluate
  the first component $h_1$ of its Laurent expansion in powers of $z^{-1}$.\\
\end{description}
The function
\begin{equation}
  \label{eq:4.45}
  u(x,t_3,\dots)=2 h_1\vert_{t_1=x}
\end{equation}
is then a solution of the KdV equation.

\subsection{The relation with the Sato approach}

The equations \rref{eq:4.27} make sense for an arbitrary Laurent series $h$ of 
the form \rref{eq:4.16}, even if it is not a solution of the Riccati equation 
$h_x+h^2=u+z^2$. Hence they define, for every $j$, a system of PDEs for the 
coefficients $h_l$. We will show\footnote{See also the papers 
\cite{Ch78,Wil81}.} that these systems are equivalent to the celebrated KP 
hierarchy of the Kyoto school (see the lectures by Satsuma in these volume). 
The usual definition of the KP equations can be summarized
as follows.
Let $\Psi{\CD}$ be the ring of pseudodifferential operators on the circle. 
It contains as a subring the space $\CD$ of {\em purely differential} 
operators. Let us denote with $\left(\cdot\right)_+$ the natural projection
from $\Psi{\CD}$ onto $\CD$.
Let $Q$ be a monic operators of degree $1$,
\begin{equation}
Q=\del-\sum_{j\ge 1} q_j \del^{-j}\ .
\end{equation}
The KP hierarchy is the set of Lax equations for $Q$
\begin{equation}\label{satokp}
\dpt{}{j}Q = [\left(Q^j\right)_+,Q]\ .
\end{equation}
The aim of this subsection is to show that such a Lax 
representation just arises as a kind of a Euler form of 
the equations~\rref{eq:4.27}. Before stating the next result, we must observe 
that the relations \rref{eq:4.28} can be solved backwards, in such a way to 
define the Fa\`a di Bruno elements $\h{j}$ for all $j\in\mathbb Z$.
\begin{prop}
Suppose the series $h$ of the form \rref{eq:4.16} to evolve 
according to a conservation law, 
\begin{equation}\label{conlaw}
\dpt{h}{}=\del_x H,
\end{equation} 
for an arbitrary $H$. Then the Fa\`a di Bruno elements $\h{j}$, 
for $j\in\mathbb Z$, evolve according to 
\begin{equation}\label{5.1}
\left(\dpt{}{}+H\right)\h{j}=
\sum_{k=0}^{\infty}\binomial{j}{k} (\del_x^k H) \h{j-k}\ ,
\end{equation}
where 
\[
\binomial{j}{k}=\frac{j (j-1)\cdots (j-k+1)}{k!}\ ,\qquad \binomial{j}{0}=1\ .
\]
\end{prop}
Now we consider the map $\phi:\CL\to {\Psi}{\CD}$, from 
the space of Laurent series to the ring of pseudodifferential 
operators on the circle, which acts on the Fa\`a di Bruno basis 
according to 
\begin{equation}
\phi(\h{j})=\del^j.\label{5.5}
\end{equation}
This map is then extended by linearity (with respect to multiplication
by a {\em function} of $x$) to the whole space $\CL$.
\begin{defi} We call Lax operator of the KP theory the 
image 
\begin{equation}
Q=\phi(z)\label{5.6}
\end{equation}
of the first element of the standard basis in $\CL$.
\end{defi}
If the $q_j$ are the components of the expansion of $z$ on the 
Fa\`a di Bruno basis,
\begin{equation}
z=\h{1}-\sum_{j\geq 1}q_jh^{(-j)}\ ,\label{5.7}
\end{equation}
then we can write
\begin{equation}
Q=\del-\sum_{j\ge 1} q_j \del^{-j}
\label{5.8}
\end{equation}
according to the definition of the map $\phi$. We note that 
equation \rref{5.7} uniquely defines the coefficients $q_j$ as 
differential polynomials of the components $h_j$ of $h(z)$:
\begin{equation}
\begin{array}{l}
q_1=h_1,\quad
q_2=h_2,\quad
q_3=h_3+h_1^2\\
q_4=h_4+3h_1h_2-h_1h_{1x}\\
\dots \dots
\end{array}\label{5.9}
\end{equation}
This is an invertible relation between the 
$h_j$ and the $q_j$, so that equation \rref{5.7} may be seen as a change 
of coordinates in the space $\CL$.
\begin{prop}
The map $\phi$ has the following three properties: 
\begin{description}
\item[i)]
Multiplying a vector of the Fa\`a di 
Bruno basis by a power $z^k$ of $z$ yields
\begin{equation}
\phi(z^k\cdot \h{j})={\del }^j\cdot Q^k\ .\label{5.10}
\end{equation}
\item[ii)]
The evolution along a conservation law of the form
\[
\dpt{h}{}=\del_x\left( \sum_{k}H_kz^k\right)
\]
translates into
\begin{equation}
\dpt{}{}\left(\phi(\h{j})\right)=\sum_{k}[{\del}^j,H_k]\cdot Q^k\ .\label{5.12}
\end{equation}
\item[iii)]
If $\pi_+$ and $\Pi_+$ are respectively the projection onto
the positive part $\CH_+\subset \CL$ and ${\CD}\subset\Psi{\CD}$, then
\begin{equation}
\phi\circ \pi_+=\Pi_+\circ \phi\ .\label{5.16}
\end{equation}
\end{description}
\end{prop}
To obtain the Sato form of the equations~\rref{eq:4.27}, 
we derive the equation
\begin{equation}
z=\h{1}-\sum_{l\geq 1}q_l\h{-l}\label{5.17}
\end{equation}
with respect to the time $t_j$, getting 
\begin{equation}
\sum_{l\geq 1}\frac{\del q_l}{\del t_j}\h{-l}=
\frac{\del \h{1}}{\del t_j}- \sum_{l\geq 1}q_l\frac{\del \h{-l}}{\del t_j}\ .
\label{5.18}
\end{equation}
Applying the map $\phi$ to both sides of this equation
we obtain 
\begin{equation}
\sum_{l\geq 1}\frac{\del q_l}{\del t_j}{\del }^{-l}=
\sum_{k\geq 1}[\del ,H^j_k]Q^{-k}-
\sum_{k\geq 1}q_l[\del^{-l} ,H^j_k]Q^{-k},
\label{5.19}
\end{equation}
or
\begin{equation}
\frac{\del Q}{\del t_j}+\sum_{k\geq 1}[Q,H^j_k]Q^{-k}=0\ .\label{5.20}
\end{equation}
Finally, we introduce the operator
\begin{equation}
B^{(j)}=\phi(\Ha{j})=\phi\left(z^j+\sum_{k\geq 1}H^j_kz^{-k}\right)=
Q^j+\sum_{k\geq 1}H^j_kQ^{-k}\label{5.21}
\end{equation}  
associated with the current density $\Ha{j}$, and we note that
\begin{equation}
B^{(j)}=\phi(\pi_+(z^j)))=\left(\phi(z^j)\right)_+
=\left(Q^j\right)_+\ .\label{5.22}
\end{equation}
Thus we can write~\rref{5.20} in the final form
\begin{equation}
\frac{\del Q}{\del t_j}+[Q,\left(Q^j\right)_+]=0\ ,\label{5.23}
\end{equation}
which coincides with equation \rref{satokp}.
\newpage

\section{Lax representation of the reduced KdV flows}\label{lect:5}
In this lecture we want to investigate more accurately the properties of the
stationary KdV flows, that is, of the equations induced by the KdV \ger y on
the finite--dimensional invariant submanifolds of the singular points of any
equation of the \ger y. Examples of these reductions have already been
discussed in the first lecture. In the third lecture we realized, in a
couple of examples, that the reduced flows were still \bih. Although not at
all surprising, this property is somewhat mysterious, since it is not yet well 
understood how the Poisson pairs of the reductions are related to the original 
Poisson pairs of the KdV equation.
Moreover, even if the subject is quite old and classical
(see, e.g., \cite{BoNo,DKN,BuEnLe}), it was still lacking 
in the literature  a systematic and coordinate free proof that such reduced 
flows are \bih\ (see, however, \cite{AFW,To95}).
In this lecture we will not provide such a proof, which is contained
in~\cite{fmpz2}, but we will give a sufficiently systematic algorithm to
compute the reduced Poisson pair. This algorithm is based on the study of the
Lax representation of the reduced equations.
\subsection{Lax representation}\label{lect:51}
In this section we associate a Lax matrix (polynomially depending on $\la$)
with each element $\Ha{j}$. This matrix naturally arises from a change of
basis in the linear space $\CH_+$ attached to  the point $h$. So far we have
introduced two bases:
\begin{description}
\item[i)] The moving frame $\{\ha{j}\}$;
\item[ii)] The canonical basis $\{\Ha{j}\}$.
\end{description}
Presently we introduce a third basis by exploiting the constraint
\begin{equation}
  \label{eq:5.1}
  \la(\CH_+)\subset \CH_+\ ,
\end{equation}
characteristic of the KdV theory. The new basis is formed by the multiples
$\{\la^j\Ha{0},\la^j\Ha{1}\}$ of the first two currents. Formally we 
define
\begin{displaymath}
  \mbox{ iii) the Lax basis: } (\la^j,\la^j h)\ .
\end{displaymath}
The use of this basis leads to a new representation of the currents $\Ha{j}$,
where each current is written as a linear combination of the first two,
$\Ha{0}=1$ and $\Ha{1}=h$, with coefficients that are polynomials in
$\la$. Let us consider a few examples:
\begin{equation}
  \label{eq:5.2}
  \begin{split}
\Ha{0}&=1+ 0\cdot h\\
\Ha{1}&=0\cdot 1+ 1\cdot h\\
\Ha{2}&=\la\cdot 1+ 0\cdot h\\
\Ha{3}&=-h_2\cdot 1+ (\la-h_1)\cdot h\\
\Ha{4}&=\la^2\cdot 1\\
\Ha{5}&=(-\la h_2+h_1h_2-h_4)\cdot 1+ (\la^2-\la h_1+h_1^2-h_3)\cdot h\>\ .
\end{split}
\end{equation}
This new representation also affects our way of writing the action of the 
operators $\big(
\Bdpt{}{j}+\Ha{j}\big)$. Let these operators act on $\Ha{0}$ and $\Ha{1}$. For 
the basic invariance condition~\rref{eq:4.34}, we get an element in $\CH_+$
which can be represented on the Lax basis. As a result we can write 
\begin{equation}
\label{eq:5.3}
\left(\Bdpt{}{j}+\Ha{j}\right) \vec2{1}{h}=L^{(j)}(\la)\vec2{1}{h}\ ,
\end{equation}
where $L^{(j)}(\la)$ is the {\em Lax matrix} associated with the current
$\Ha{j}$. We shall see below the explicit form of some of these matrices. 

It becomes now very easy to rewrite the Central System in the form of 
equations on the Lax matrices $\Lax{j}(\la)$. We simply have to notice that
the equations \rref{eq:4.35} entail the ``exactness condition''
\begin{equation}
  \label{eq:5.4}
  \Bdpt{\Ha{j}}{k}= \Bdpt{\Ha{k}}{j}\ ,
\end{equation}
from which it follows that the operators $\big(\Bdpt{}{j}+\Ha{j}\big)$ and 
$\big(
\Bdpt{}{k}+\Ha{k}\big)$ commute:
\begin{equation}
  \label{eq:5.5}
  \left[ \Bdpt{}{j}+\Ha{j}, \Bdpt{}{k}+\Ha{k}\right]=0\>\ .
\end{equation}
It is now sufficient to evaluate this condition on $(\Ha{0},\Ha{1})$ and
to expand on the Lax basis to find the ``zero curvature representation'' of the
KdV \ger y:
\begin{equation}
  \label{eq:5.6}
  \Bdpt{\Lax{j}}{k}-\Bdpt{\Lax{k}}{j}+\big[\Lax{j},\Lax{k}\big]=0\>\ .
\end{equation}
Suppose now that we are on the invariant submanifold formed by  the singular
points of the $j$--th member of the KdV \ger y. On this submanifold
\begin{equation}
  \label{eq:5.7}
  \Bdpt{\Lax{k}}{j}=0\qquad\forall\, k,
\end{equation}
and the zero curvature representation becomes the Lax representation
\begin{equation}
  \label{eq:5.8}
\Bdpt{\Lax{j}}{k}=\big[\Lax{k},\Lax{j}\big]\>\ .  
\end{equation}
We have thus shown that all the stationary reductions of the KdV \ger y admit
a Lax representation. As a matter of fact, this Lax representation
coincides~\cite{fmpz2} with the Lax representation of the GZ
systems on Lie--Poisson manifolds studied in Section 3. The latter are \bih\ 
systems. Therefore, we
end up stating that the stationary reductions of the KdV theory are \bih, and
we can construct the associated Poisson pairs. We shall now see a couple of
examples.

\subsection{First example}\label{lect:52}
We study anew the simplest invariant submanifold  of the KdV \ger y, defined
by the equation
\begin{equation}
  \label{eq:5.9}
  u_{xxx}-6u u_x =0 \ .
\end{equation}
In this example we consider the constraint from the point of view of the
Central System. Since the constraint is the stationarity of the time $t_3$, we 
have to consider only the first three Lax matrices.
As for the matrix $\Lax{1}$, the following computation,
\begin{equation}
  \label{eq:5.10}
  \begin{split}
\big(\Bdpt{}{1}+\Ha{1}\big) &1=0\cdot 1+1\cdot h\\
\big(\Bdpt{}{1}+\Ha{1}\big)&\Ha{1}\eqcon{4.35}
\Ha{2}+2h_1 \eqcon{5.2}(\la+2h_1)\cdot 1 +0\cdot h\ ,\\
\end{split}
\end{equation}
shows that 
\begin{equation}
  \label{eq:5.11}
  \Lax{1}=\left (\begin {array}{cc} 0&1
\\ \lambda+2\,h_{{1}}&0\end {array}\right )\ .
\end{equation}
Similarly, the computation
\begin{equation}  \label{eq:5.12}
  \begin{split}
\big(\Bdpt{}{3}+\Ha{3}\big) 1&\eqcon{5.2}-h_2\cdot 1+ (\la-h_1)\cdot h\\
\big(\Bdpt{}{3}+\Ha{3}\big) \Ha{1}&\eqcon{4.35}
\Ha{4}+h_1\Ha{2}+h_2\Ha{1}+h_3+H^3_1 \\ &\eqcon{5.2}
(\la^2+\la h_1+2h_3-h_1^2)\cdot 1 +h_2\cdot h\\
\end{split}
\end{equation}
yields
\begin{equation}
  \label{eq:5.13}
\Lax{3}=  \left (
\begin {array}{cc} -h_{{2}}&\lambda-h_{{1}}
\\{\lambda}^{2}+h_{{1}}\lambda+2h_{{3}}-{h_1}^2 &h_2 
\end{array} \right).
\end{equation}
On the submanifold $M_3$ defined by
equation \rref{eq:5.9} this matrix verifies 
the
Lax equation
\begin{equation}
  \label{eq:5.14}
  \Bdpt{\Lax{3}}{1}=\big[\Lax{1},\Lax{3}\big]\>\ .  
\end{equation}
This equation completely defines the time evolution of the first three
components $(h_1,h_2,h_3)$ of the current $\Ha{1}=h$. These components play the 
role of coordinates on $M_3$. We get
\begin{equation}
  \label{eq:5.15}
  \begin{split}
\Bdpt{h_1}1&=-2h_2\\
\Bdpt{h_2}1&=-2h_3-{h_1}^{2}\\
\Bdpt{h_3}1&=-4h_1h_2\\
\end{split}
\end{equation}
By the change of coordinates 
\begin{displaymath}
h_1=\frac12 u\ ,\qquad h_2=-\frac14u_x\ ,\qquad h_3=\frac18(u_{xx}-u^2)\ ,
\end{displaymath}
coming from the inversion~\rref{eq:4.15} of the
Miura map, these equations take the form
\begin{equation}
  \label{eq:5.16}
   \Bdpt{u}{1}=u_x,\quad  \Bdpt{u_x}{1}=u_{xx},\quad  
\Bdpt{u_{xx}}{1}=6uu_x\ , 
\end{equation}
already encountered in Lecture~\ref{lect:1}. This shows explicitly the
connection between the two points of view. 

To find the connection between these equations and the GZ equations dealt with
in the first example of Lecture~\ref{lect:3}, we compare the Lax matrix
\begin{displaymath}
  \Lax{3}(\la)=\la^2\mat2{0}{0}{1}{0}+\la\mat2{0}{1}{h_1}{0}+
\mat2{-h_2}{-h_1}{2h_3-h_1^2}{h_2}
\end{displaymath}
with the Lax matrix
\begin{displaymath}
  S(\la)=\la^2\mat2{0}{0}{1}{0}+\la\mat2{p_1}{1}{q_1}{-p_1}+
\mat2{p_0}{-(q_1+p_1^2)}{q_0}{-p_0}
\end{displaymath}
associated with the points of the symplectic leaf defined by
\rref{eq:3.18}. We easily identify $\Lax{3}$ with the restriction of
$S(\la)$ to $p_1=0$ upon setting
\begin{equation}
  \label{eq:5.17}
  p_0=-h_2,\quad q_1=h_1,\quad q_0=2h_3-h_1^2\>\ .
\end{equation}
By comparing these equations with the projection~\rref{eq:3.21}, which allows
to pass from the symplectic leaf $S$  to the quotient space $N=S/E$, we
obtain the change of coordinates
\begin{equation}
  \label{eq:5.18}
  u_1=h_1,\quad u_2=-h_2,\quad u_3=2h_3-h_1^2,
\end{equation}
connecting the reduction~\rref{eq:5.15} of the Central System to the GZ
system~\rref{eq:3.29} dealt with in the third Lecture. The latter was, by
construction, a \bih\ system.  We argue that also the reduction of the Central 
System herewith considered is a \bih\ \vefi, and that its Poisson pair is
obtained by geometric reduction. Basic for this identification is the property 
of the Lax matrix $\Lax{3}$ of being a {\em section} of the fiber bundle
$\pi:S\to S/E$ appearing in the geometric reduction. It is this property
which allows to set an invertible relation among  the coordinates
$(u_1,u_2,u_3)$, coming from the geometric reduction, and the coordinates
$(h_1,h_2,h_3)$ coming from the reduction of the Central System.
\subsection{The generic stationary submanifold}\label{lect:53} 
It is now not hard to give the general form of the matrices $\Lax{j}$ for an
arbitrary odd integer $2j+1$. First we observe that
\begin{equation}
  \label{eq:5.19}
  \big(\Bdpt{}{2j+1}+\Ha{2j+1}\big)
  1=\Ha{2j+1}\eqcon{\ref{eq:4.26}}-\frac12\ala{j}_x+\ala{j}h\>\ .
  \end{equation}
Then we notice that
\begin{equation}  \label{eq:5.20}
\begin{split}
  &\big(\Bdpt{}{j}+\Ha{j}\big)h=\Ha{j+1}+\sum_{l=1}^jh_l\Ha{j-l}+ H^j_1\\
&\qquad =-\frac12\big(\ala{j+1}_x+\sum_{l=1}^jh_l \ala{j-1}_x\big)+H^j_1
+\big(\ala{j+1}+\sum_{l=1}^jh_l\ala{j-1}\big)h\ .\\
\end{split}
\end{equation}
Therefore
\begin{equation}
  \label{eq:5.21}
  \Lax{j}=\mat2{-\frac12\ala{j}_x}{\ala{j}}
{-\frac12(\ala{j+1}_x+\sum_{l=1}^jh_l\ala{j-1}_x)+H^j_1}
{\ala{j+1}+\sum_{l=1}^j
h_l\ala{j-1}}\ .
\end{equation}
By using the definition \rref{eq:4.21} of the Lenard series $\al(z)$ of which 
the polynomials $\ala{j}$ are the partial sums,
it is easy to prove that $\Lax{j}$ is a traceless matrix.

We leave to the reader to specialize the matrix $\Lax{5}$, and to write
explicitly the Lax equations
\begin{equation}
  \label{eq:5.22}
   \Bdpt{\Lax{5}}{1}=\big[\Lax{1},\Lax{5}\big]\ ,\qquad  
\Bdpt{\Lax{5}}{3}=\big[\Lax{3},\Lax{5}\big]\>\ .    
\end{equation}
They should be compared with the reduced KdV equations~\rref{eq:1.30}
and~\rref{eq:1.31} on the invariant submanifold defined by the constraint
\begin{equation}
  \label{eq:5.23}
  u_{xxxxx}-10u u_{xxx}-20u_x u_{xx}+30u^2u_x=0\>\ .
\end{equation}
They should also be compared with the GZ equations obtained via the geometric
reduction process applied to the Lie--Poisson pairs defined on three copies of
$\fraksl(2)$ by equations \rref{eq:3.31} and~\rref{eq:3.32}. We have not
displayed explicitly these equations yet. We will give their form in the next
lecture.
\subsection{What more?}\label{lect:54} 
There is nothing ``sacred'' with the KdV theory. As we know, it is related
with the constraint
\begin{equation}
  \label{eq:5.24}
  z^2(\CH_+)\subset (\CH_+)\ ,
\end{equation}
which defines an invariant submanifold of the Central System. Many other
constraints can be considered. For instance, the constraint
\begin{equation}
  \label{eq:5.25}
  z^3(\CH_+)\subset (\CH_+)
\end{equation}
leads to the so--called Boussinesq theory, and is studied
in~\cite{fmt}. What is remarkable is that the change of constraint
does not affect the algorithm for the study of the reduced equations. All the
previous reasonings are valid without almost no change. The only difference
resides in the fact that the computations become more involved. This remark 
allows 
to better appreciate the meaning of the process leading from the KdV equation
to the Central System. We have not only given a new formulation to known
equations. We have actually found a much bigger \ger y, possessing remarkable
properties, which coincides with the KdV \ger y on a (small) proper invariant
subset. The integrability properties belong to the bigger \ger y, and hold 
outside the KdV submanifold. Many other interesting equations can be found by 
other
processes of reduction. There is some evidence that a very large class
of evolution equations possessing
some integrability properties can be eventually recovered as a suitable
reduction of the Central System, or of strictly related systems.
However, we shall not
pursue this point of view further, since it would lead us too far away from
our next topic, the {\em separability\/} of the reduced KdV flows.
\newpage 
\section{Darboux--Nijenhuis coordinates and Separability}\label{lect:6}
In this lecture we shall consider the reduced KdV flows from a different point 
of view. Our aim is to probe the study of the geometry of the Poisson pair
which, as realized in the third and fifth lectures, is associated with these
flows. The final goal is to show the existence of a suitable set of
coordinates defined by and adapted to the Poisson pair. They are called {\em
Darboux--\Nij} coordinates. We shall prove that they are
separation coordinates for the Hamilton--Jacobi equations associated with the
reduced flows.

To keep the presentation within a reasonable size, we shall
mainly deal with a particular example, and we shall not discuss
thoroughly the theoretical background, referring to \cite{fmpz2} for more 
details. We shall use the example to display the 
characteristic features of the geometry of the reduced manifolds. The reader
is asked to believe that all that will be shown is general inside the class of 
the reduced stationary KdV manifolds, whose Poisson pencils are of maximal
rank. A certain care must be used in trying to extend these conclusions to
other examples like the Boussinesq stationary reductions, 
whose Poisson pencils are not of maximal rank. 
They will not be covered in these lecture notes.
The example worked out is the reduction of the first and the third KdV
equations 
on the invariant submanifold defined by the equation
\begin{equation}
  \label{eq:6.1}
   u_{xxxxx}-10u u_{xxx}-20u_x u_{xx}+30u^2u_x=0\ ,
\end{equation}
a problem addressed at the end of Section~\ref{lect:53}. 
\subsection{The Poisson pair}\label{lect:61}
As we mentioned several times, the invariant submanifold $M_5$ defined by
equation \rref{eq:6.1} has dimension five. From the standpoint of the Central
System, it is characterized by the two equations
\begin{equation}
  \label{eq:6.2}
  z^2(\CH_+)\subset (\CH_+)\ ,\qquad 
\Ha{5}h=\la^3+\sum_{l=1}^5 h_l\Ha{5-l}+H^5_1\>\ .
\end{equation}
We recall that the first constraint means that, inside the big cell of the
Sato Grassmannian, we are working on the special submanifold corresponding to
the KdV theory. The second constraint means that, inside the phase 
space of the KdV theory, we are working on the set~\rref{eq:6.1} of 
singular points of the fifth flow. The two constraints play the following
roles. The first constraint sets up a relation among the currents $\Ha{j}$:
All the currents are expressed as linear combinations (with polynomial
coefficients) of the first two currents $\Ha{0}=1$ and $\Ha{1}=h$. So this
constraint drastically reduces the number of the unknowns $H^j_l$ to the
coefficients $h_l$ of $h$. The second constraint then further cuts the
degrees of freedom to a finite number, by setting relations among the
coefficients $h_l$. It can be shown that only the first five  coefficients
$(h_1,h_2,h_3,h_4,h_5)$ survive as free parameters. All the other coefficients 
can 
be expressed as polynomial functions of the previous ones. By a process of
elimination of the exceeding coordinate, 
one proves that the restriction of the first and third flows of the KdV \ger y 
are represented by the following differential equations:
\begin{equation}
\label{eq:6.4}
\begin{split}
&\Bdpt{h_1}1=-2h_2  \\
&\Bdpt{h_2}1=-2h_3-{h_1}^{2}\\  
&\Bdpt{h_3}1=-2h_1h_2-2h_4  \\
&\Bdpt{h_4}1=-2h_5-{h_2}^{2}-2h_1h_3 \\
&\Bdpt{h_5}1=-4h_3h_2+2{h_1}^{2}h_2-4h_1h_4\\  
\end{split}
\end{equation}
and
\begin{equation}
  \label{eq:6.5}
\begin{split}
&\Bdpt{h_1}3= -2h_4+2h_1h_2   \\     
&\Bdpt{h_2}3= -2h_5+{h_2}^{2}+{h_1}^{3}   \\ 
&\Bdpt{h_3}3= -2h_1h_4+4{h_1}^{2}h_2-2h_3h_2    \\ 
&\Bdpt{h_4}3=-2{h_3}^{2}-2h_2h_4 +2h_1{h_2}^{2}+{h_1}^{4}+{h_1}^{2}h_3    \\
&\Bdpt{h_5}3=2{h_1}^{2}h_4-4h_3h_4+2{h_1}^{3}h_2\\
\end{split}
\end{equation}
They can also be seen as the Lax equations~\rref{eq:5.22}. However, for our
purposes, it is more important to recognize that the above equations are the
GZ equations of the Poisson pencil defined on $M_5$. This pencil can be
computed according to the reduction procedure explained in the third
lecture. The final outcome is that the reduced Poisson bivector is given by
\begin{equation}
  \label{eq:6.6}
  \begin{split}
\dot{h_1}&=2\ddd{H}{h_2}+2(h_1-\la)\Hdd{4}+2h_2\Hdd{5}\\
\dot{h_2}&=-2\Hdd{1}+2(\la-2h_1)\Hdd{3}-2h_2\Hdd{4}
+(4\la h_1-2h_3-h_1^2)\Hdd{5}\\
\dot{h_3}&=2(2h_1-\la)\Hdd{2}+(2h_3+2h_1^2-4\la h_1)\Hdd{4}+
2(h_4+h_1h_2)\Hdd{5}\\
\dot{h_4}&=2(\la-h_1)\Hdd{1}+2h_2\Hdd{2}-(2h_3+2h_1^2-4\la h_1)\Hdd{3}\\
&\qquad +(2h_5-6h_1h_3+h_2^2+2h_1^3+4\la h_3+2\la h_1^2)\Hdd{5}\\
\dot{h_5}&=-2h_2\Hdd{1}+(2h_3+h_1^2-4\la h_1)\Hdd{2}-2(h_4+h_1h_2)\Hdd{3}\\
&\qquad -(2h_5-6h_1h_3+h_2^2+2h_1^3+4\la h_3+2\la h_1^2)\Hdd{4}\>.\\
\end{split}
\end{equation}
The Casimir function of this pencil is a quadratic polynomial,
\begin{equation}
  \label{eq:6.7}
  C(\la)=C_0\la^2+C_1\la+C_2\ ,
\end{equation}
and the coefficients are
\begin{equation}
  \label{eq:6.8}
  \begin{array}{l}
C_0={h_1}^{3}-2h_1h_3+h_5\\
C_1=h_2h_4-h_1h_5+\frac32{h_1}^{2}h_3-\frac12h_1{h_2}^{2}-\frac12{h_3}^{2}-
\frac12{h_1}^{4}\\
C_2=\frac12h_3{h_2}^{2}-h_3h_5+\frac12{h_1}^{5}+h_1{h_3}^{2}-h_1h_2h_4-
\frac32{h_1}^{3}h_3+{h_1}^{2}h_5+\frac12{h_4}^{2}
\end{array}
\end{equation}
The Lenard chain is
\begin{equation}
  \label{eq:6.9}
  \begin{array}{lcl}
P_0 d C_0&=&0\\
P_0 d C_1&=&P_1 d C_0=\Bdpt{\mathbf h}{1}\\
P_0 d C_2&=&P_1 d C_1=\Bdpt{\mathbf h}{3}\\
&&P_1 d C_2=\>\>0\ ,
\end{array}
\end{equation}
where $\mathbf h$ is the vector $(h_1,h_2,h_3,h_4,h_5)$. It shows that the
reduced flows are \bih. Finally, if one uses the coordinate 
change~\rref{eq:4.17} 
from the coordinates $(h_1,h_2,h_3,h_4,h_5)$ to the coordinates
$(u,u_x,u_{xx},u_{xxx},u_{xxxx})$, one can put the equations \rref{eq:6.4}
and~\rref{eq:6.5} in the form~\rref{eq:1.30} and \rref{eq:1.31} considered in 
the first lecture.
\subsection{Passing to a symplectic leaf}\label{lect:62}
We aim to solve equations \rref{eq:6.4} and~\rref{eq:6.5} by the \HJ method. 
This
requires to set the study of such equations on a symplectic manifold. This
can be easily accomplished by noticing that these vector fields are already
tangent to the submanifold $S_4$ defined by the equation 
\begin{equation}
  \label{eq:6.10}
  C_0=E\ , 
\end{equation}
for a constant $E$.
We know that this submanifold is symplectic since $C_0$ is the Casimir 
of $P_0$. The dimension of $S_4$ is four, and the variables
$(h_1,h_2,h_3,h_4)$ play the role of coordinates on it.

For our purposes it is crucial to remark an additional property
of $S_4$: It is a \varb. This means that also the second bivector $P_1$
induces, by a process of reduction, a Poisson structure on $S_4$ compatible
with the natural restriction of $P_0$.
This is not a general situation. It holds as a consequence of a
peculiarity of the Poisson pencil~\rref{eq:6.6}. The property we are
mentioning concerns the \vefi
\begin{equation}
  \label{eq:6.11}
  Z=\ddd{}{h_5}.
\end{equation}
One can easily check that:
\begin{description}
\item[i)] $Z$ is transversal to the symplectic leaf $S_4$.
\item[ii)] The functions which are invariant along $Z$ form a Poisson
  subalgebra with respect to the pencil.
\end{description}
In simpler terms, the Poisson bracket of functions which are independent of
$h_5$ is independent on $h_5$ as well. Since they coincide with the functions
on $S_4$ (by the transversality condition), this property allows us to define a
pair of Poisson brackets also on $S_4$. The first bracket is associated with
the symplectic 2--form $\omega_0$ on $S_4$. It can be easily checked that
\begin{equation}
  \label{eq:6.12}
  \omega_0=h_1 dh_1\wedge dh_2+\frac12(dh_2\wedge dh_4+dh_5\wedge dh_1)\ .
\end{equation}
The second Poisson bracket can be represented in the form
\begin{equation}
  \label{eq:6.13}
  \parpo{f}{g}_1=\omega_0(N X_f,X_g)\ ,
\end{equation}
where $X_f$ and $X_g$ are the \ham\ \vefi s associated with the functions $f$
and $g$ by the symplectic 2--form $\omega_0$, and $N$ is a $(1,1)$--tensor
field on $S_4$, called the \Nij\ tensor associated with the
pencil (see, e.g., \cite{KM}). In our example one obtains
\begin{equation}
  \label{eq:6.14}
  \begin{split}
N=&
\big(-h_1\ddd{}{h_1}-h_2\ddd{}{h_2}+(h_3-3h_1^2)\ddd{}{h_3}-
2h_1h_2\ddd{}{h_4}\big)\otimes 
dh_1\\
&+(h_3-h_1^2)\ddd{}{h_4}\otimes dh_2+\big(\ddd{}{h_1}+2h_1\ddd{}{h_3}
+h_2\ddd{}{h_4}\big)\otimes dh_3\\ 
&+\big(\ddd{}{h_2}+h_1\ddd{}{h_4}\big)\otimes dh_4\ .\\
\end{split}
\end{equation}
Thus we arrive at the following picture of the GZ \ger y considered in this
lecture. It is formed by a pair of \vefi s, $X_1$ and $X_3$, defined by
\rref{eq:6.4} and~\rref{eq:6.5}. They are tangent to the symplectic 
leaf $(S_4,\omega_0)$ defined by equations \rref{eq:6.10} and~\rref{eq:6.12}. 
This
symplectic manifold is still \bih, and therefore there exists a \Nij\ tensor
field $N$, defined by equation \rref{eq:6.13}. The vector fields $X_1$ and $X_3$
span a Lagrangian subspace 
which is invariant 
with respect to $N$. One finds that
they obey the following ``modified Lenard recursion relations''
\begin{equation}
  \label{eq:6.15}
  \begin{array}{ccccc}
N X_1&=&X_3&+&(\mbox{Tr }N)X_1\\
N X_3&=&   &+&(-\mbox{det }N)X_1\ .
\end{array}
\end{equation}
From them we can extract the matrix 
\begin{equation}
  \label{eq:6.16}
  \Ff=\mat2{\mbox{Tr }N}{1}{-\mbox{det }N}{0}
\end{equation}
which represents the action of $N$ on the abovementioned Lagrangian subspace. 
It will play a fundamental role in the upcoming discussion of the separability 
of the \vefi s.
\begin{exer} Compute the expression of the reduced pencil on $S_4$ and check
  the form of the \Nij\ tensor, as well as the  modified Lenard recursion
  relations~\rref{eq:6.15}. 
\end{exer}
\subsection{Darboux--\Nij\ coordinates}\label{lect:63}
We are now in a position to introduce the basic tool of the theory of
separability in the \bih\ framework: The concept of Darboux--\Nij\ 
coordinates on a symplectic \varb, like $S_4$.

Given a symplectic 2--form $\omega_0$ and a \Nij\  tensor $N$ coming from a
Poisson pencil defined on a $2n$--dimensional manifold $\CM$, under the
assumption that the eigenvalues of $N$ are real and functionally independent,
one proves~\cite{Ma90}
the existence of a system of
coordinates $(\la_1,\ldots, \la_n;\mu_1,\ldots, \mu_n)$ which are canonical
for $\omega_0$,
\begin{equation}
  \label{eq:6.17}
  \omega_0=\sum_{i=1}^n d\mu_1\wedge d\la_i\ ,
\end{equation}
and which allows to put $N^*$ (the adjoint of $N$) in diagonal form:
\begin{equation}
  \label{eq:6.18}
  N^* d\la_i=\la_i d\la_i\ ,\qquad N^* d\mu_i=\la_i d\mu_i\>\ .
\end{equation}
The coordinates $\la_i$ are the eigenvalues of $N^*$, and therefore can be
computed as the {\em zeroes} of the minimal polynomial of $N$:
\begin{equation}
  \label{eq:6.19}
  \la^n+c_1\la^{n-1}+\cdots+c_n=0\ .
\end{equation}
The coordinates $\mu_j$ can be computed as the {\em values} that a conjugate
polynomial
\begin{equation}
  \label{eq:6.20}
  \mu=f_1\la^{n-1}+\cdots+f_n
\end{equation}
assumes on the eigenvalues $\la_j$, that is, 
\begin{equation}
  \label{eq:6.21}
  \mu_j=f_1\la^{n-1}_j+\cdots+f_n,\quad j=1,\ldots, n\>\ .
\end{equation}
The determination of this polynomial, which is not uniquely defined by the
geometric structures present in  the theory,  requires a certain
care. Although there is presently a sufficiently developed theory on the 
Darboux--\Nij\ 
coordinates and on their computation, for the sake of brevity we shall not
tackle this problem, but rather limit ourselves to display these polynomials
in the example at hand. They are
\begin{equation}
  \label{eq:6.23}
  \begin{split}
&\la^2-h_1\la+(h_1^2-h_3)=0\\
&\mu-h_2\la+(h_1h_2-h_4)=0\\
\end{split}
\end{equation}
The important idea emerging from the previous discussion is that the GZ
equations are often coupled with a special system of coordinates related with
the Poisson pair.
\begin{exer} Check that the polynomials \rref{eq:6.23} define a system of
  Darboux--\Nij\  coordinates for the pair $(\omega_0,N)$ considered above.
\end{exer}
\subsection{Separation of Variables}
We start from the classical St\"ackel theorem on the separability, in
orthogonal coordinates, of the \HJ equation associated with the natural
\ham 
\begin{equation}
  \label{eq:7.1}
  H(q,p)=\frac12\sum g^{ii}(q)p_i^2+V(q_1,\ldots, q_n)
\end{equation}
on the cotangent bundle of the configuration space. According to
St\"ackel, this \ham\ is separable if and only if there exists as invertible
matrix $S(q_1,\ldots, q_n)$ and a vector $U(q_1,\ldots, q_n)$ such that $H$ is 
among the solutions $(H_1,\ldots, H_n)$ of the linear system
\begin{equation}
  \label{eq:7.2}
  \frac12 p_i^2=U_i(q)+\sum_{j=1}^n S_{ij}(q) H_j\ ,
\end{equation}
and $S$ and $U$ verify the St\"ackel condition:
\begin{center}{\em 
The rows of $S$ and $U$ depend only on the corresponding coordinate.}
\end{center} 
This
means for instance that the elements $S_{1j}$ and $U_1$ depend only on the
first coordinate $q_1$, and so on. Such a matrix $S$ is called a  St\"ackel
matrix (and $U$ a St\"ackel vector). 

The strategy we shall follow to prove the separability of the \HJ\ equations
associated with the GZ \vefi s $X_1$ and $X_3$ on the manifold $S_4$ considered 
above, is to show that the Darboux--\Nij\ coordinates allow
to define a St\"ackel matrix for the corresponding \ham s.

The construction of the  St\"ackel matrix starts from the matrix $\Ff$ which
relates the \vefi\ $X_1$ and $X_3$ to the \Nij\ tensor $N$ (see
equation \rref{eq:6.16}). One can prove that this matrix satisfies the 
remarkable identity
\begin{equation}
  \label{eq:7.3}
  N^* d\Ff=\Ff d\Ff\>\ .
\end{equation}
This is a matrix equation which must be interpreted as follows: $d \Ff$ is a
matrix of 1--forms, and $N^*$ acts separately on each entry of this
matrix; $\Ff d\Ff$ denotes the matrix multiplication of the matrices $\Ff$ and 
$d \Ff$, which amounts to linearly combine the 1--forms appearing in $d
\Ff$. In our example, equation \rref{eq:7.3} becomes
\begin{equation}
  \label{eq:7.4}
  \begin{array}{ccccc}
N^* d (\mbox{Tr}N)&=&-d(\mbox{det}N)&+&(\mbox{Tr}N)\, d (\mbox{Tr}N)\\
N^* d (\mbox{det}N)&=&                &+&(\mbox{det}N)\, d (\mbox{Tr}N)
\end{array}
\end{equation}
\begin{exer} Check that this equations are verified by the \Nij\
  tensor~\rref{eq:6.14}. 
\end{exer}
We leave for a moment the particular case we are dealing with, and we suppose 
that, on a symplectic \varb\ fulfilling the conditions of Subsection 
\ref{lect:63}, a family of $n$ \vefi s $(X_{1},X_3,\dots,X_{2n-1})$ is given.
We assume that they are \ham\ with respect to $P_0$, say, $X_{2i-1}=P_0 dC_i$, 
and that there exists a matrix $\Ff$ such that 
\begin{equation}
\label{deffrob}
NX_{2i-1}=\sum_{j=1}^n \Ff_i^j X_{2j-1}\qquad \mbox{for all } i\ .
\end{equation}
Finally, we suppose that $\Ff$ satisfies condition \rref{eq:7.3}. Then, 
from the matrix $\Ff$ we build up the matrix $\Tt$ whose rows are the
left--eigenvectors of $\Ff$. In other words, we construct a matrix $\Tt$ such
that
\begin{equation}
  \label{eq:7.6}
  \Ff=\Tt^{-1}\La \Tt\ ,
\end{equation}
where $\La=\mbox{diag}(\la_1,\dots,\la_n)$ is the diagonal matrix of the
eigenvalues of $\Ff$, coinciding with the eigenvalues of $N$.
The matrix $\Tt$ is normalized by imposing that in each row there is a
constant component. A suitable normalization criterion, for instance, is to
set the entries in the last column equal to $1$. 
\begin{theorem} If the matrix $\Ff$ verifies condition~\rref{eq:7.3} (as it is 
  always true in our class of examples), then the matrix $\Tt$ is a 
(generalized) St\"ackel matrix in the Darboux--\Nij\ coordinates.
\end{theorem}

This theorem means that the rows of the matrix $\Tt$ verify the following  
generalized St\"ackel condition: The entries of the first row of $\Tt$ depend
only on the canonical pair $(\la_1,\mu_1)$, those of the second row on
$(\la_2,\mu_2)$, and so on. With respect to the classical case recalled at the
beginning of this lecture, we notice that by generalizing the class of \ham s
considered, we have been obliged to extend a little bit the notion of St\"ackel
matrix. However, this extension does not affect the theorem of
separability. Indeed, as a consequence of the fact that the matrix $\Ff$ is
defined by the \vefi s $(X_{1},X_3,\dots,X_{2n-1})$ themselves through equation 
\rref{deffrob}, one can 
prove that $\Tt$ is a St\"ackel matrix for the corresponding \ham s 
$(C_1,\dots,C_n)$. 
\begin{theorem} The column vector 
\begin{equation}\label{eq:7.7}
\mathbf U=\Tt \mathbf C\ ,
\end{equation}
where $\mathbf C$ is the column vector of the \ham s   $(C_1,\ldots, C_n)$,
verifies the (generalized) St\"ackel condition in the Darboux--\Nij\ 
coordinates. This means that the first component of $\mathbf U$ depends
only on
the pair $(\la_1,\mu_1)$, the second on $(\la_2,\mu_2)$, and so on.
\end{theorem}
We shall not prove these two theorems here, preferring to see them ``at work'' 
in the example at hand. First we consider the matrix $T$. Due to the form 
\rref{eq:6.16} of the matrix $\Ff$, it is easily proved that
\begin{equation}
  \label{eq:7.8}
  \Tt=\mat2{\la_1}{1}{\la_2}{1}\>\ .
\end{equation}
Indeed, the equation $\Tt\Ff=\La\Tt$ follows directly from the characteristic
equation for the tensor $N$. It should be noted that the matrix $\Tt$ has been 
computed without computing explicitly the eigenvalues $\la_1$ and $\la_2$. It
is enough to use the first of equations \rref{eq:6.23}, 
defining the Darboux--\Nij\ coordinates. The matrix $\Tt$ clearly possess the   
St\"ackel
property (even in the classical, restricted sense).

The vector $\mathbf U$ can be computed as well without computing explicitly
the coordinates $(\la_j,\mu_j)$. It is sufficient, once again, to use
the equations \rref{eq:6.23}. We now pass to prove that equation \rref{eq:7.7}, 
in our example, has the particular form
\begin{equation}
  \label{eq:7.9}
  \begin{split}
\frac12\mu_1^2-\frac12\la_1^2-E\la_1^2&=\la_1 C_1+C_2\\
\frac12\mu_2^2-\frac12\la_2^2-E\la_2^2&=\la_2 C_1+C_2\ .\\
\end{split}
\end{equation}
We notice that proving this statement is tantamount to proving that the
following equality between polynomials,
\begin{equation}
  \label{eq:7.10}
  \mu(\la)^2-\la^5=2 C(\la)\ ,
\end{equation}
is verified in correspondence of the eigenvalues of $N$. This can be done as
follows. Let us write the polynomials defining the Darboux--\Nij\ coordinates
in the symbolic form
\begin{equation}
  \label{eq:7.11}
  \begin{array}{l}
\la^2=e_1\la+e_2\\
\mu=f_1\la+f_2\>\ .
\end{array}
\end{equation}
The coefficients $(e_j,f_j)$ of these polynomials must be regarded as known
functions of the coordinates on the manifold. By squaring the second polynomial 
and by eliminating $\la^2$ by means of the first equation, we get
\begin{equation}
  \label{eq:7.12}
  \mu^2=f_1^2(e_1\la+e_2)+2f_1f_2\la+f_1f_2=(f_1^2e_1+2f_1f_2)\la
+(f_1^2 e_2+f_2^2)\>\ . 
\end{equation}
In the same way we obtain
\begin{equation}
  \label{eq:7.13}
\begin{split}
  \la^5=\la\cdot\la^4 &=\la[(e_1^3+2e_1e_2)\la+(e_1^2 e_2+e_2^2)]\\
&=(e_1^4+3e_1^2e_2+e_2^2)\la+(e_1^3e_2+2e_1e_2^2)\>. 
\end{split}
\end{equation}
Finally,
\begin{equation}
C(\la)=C_0\la^2+C_1\la+C_2=(C_0 e_1+C_1)\la+(C_0 e_2+C_2)\ .
\end{equation}
By inserting these expressions into equation \rref{eq:7.10}, 
we see that the resulting equation splits into two parts, according to the
``surviving'' powers of $\la$:
\begin{equation}
  \label{eq:7.15}
  \begin{split}
\la&:\quad (f_1^2e_1+2f_1f_2)-(e_1^4+3e_1^2e_2+e_2^2)=
2(C_0 e_1+C_1)\\
1&: \quad (e_1^2+e_2+e_2^2)- (e_1^3e_2+2e_1e_2^2)=2(C_0 e_2+C_2)\ .
\end{split}
\end{equation}
This method allows to reduce the proof of the separability of the \HJ\ 
equation(s) to the
procedure of checking  that explicitly known functions identically 
coincide on the manifold.
 
We end our discussion of the separability at this point. Our aim was simply to 
introduce the method of Poisson pairs, and to show by means of concrete
examples how it can be profitably used to define and solve special classes of
integrable \ham\ equations. We hope that the examples discussed in these
lectures might be successful in giving at least a feeling of the nature and
the potentialities of this method.


\newpage
\begin{thebibliography}{10}

\bibitem{Alber}
S.I. Alber, {\em On stationary problems
for equations of Korteweg--de Vries type}.
Comm. Pure Appl. Math. {\bf 34} (1981), 259--272.
 
\bibitem{AFW} M. Antonowicz, A.P. Fordy, S. Wojciechowski, {\em Integrable
stationary flows, Miura maps and bi--Hamiltonian structures.} Phys. Lett. A
{\bf 124} (1987), 455--462.

\bibitem{BoNo}
O.I. Bogoyavlensky, S.P. Novikov, 
{\em The relationship between Hamiltonian
formalism of stationary and non-stationary problems.} 
\faa{10}{1976}{8--11}.

\bibitem{BuEnLe} V.M. Buchstaber, V.Z. Enolskii, D.V. Leykin, 
{\em Hyperelliptic Kleinian functions and applications}. In: {\sl Solitons,
  Geometry and Topology: On the Crossroad}, AMS Trans. Ser. 2, Advances in
Math. Sciences (V.M. Buchstaber and S.P. Novikov, eds.), Vol. {\bf 179}, 1997,
pp.\ 1--34.

\bibitem{CMP}
P. Casati, F. Magri, M. Pedroni,
{\em Bi-Hamiltonian Manifolds and $\tau$--function.}
In: Mathematical Aspects of Classical
Field Theory 1991 (M. J. Gotay et al.\ eds.),
Contemporary Mathematics vol. {\bf 132}, 
American Mathematical Society,
Providence, R. I., 1992, pp.\ 213-234.

\bibitem{Ch78}
I.V. Cherednik, {\em Differential equations
for the Baker-Akhiezer functions of Algebraic curves.} Funct.
Anal. Appl. {\bf 10} (1978), 195--203.

\bibitem{DJKM}
E. Date, M. Jimbo, M. Kashiwara, T. Miwa,
{\em Transformation Groups for Soliton Equations.}
Proceedings of R.I.M.S. Symposium on Nonlinear IntegrableSystems--Classical 
Theory and Quantum Theory
(M. Jimbo, T. Miwa, eds.),
World Scientific, Singapore, 1983, pp.\ 39--119.

\bibitem{Di-B91}
L.A. Dickey, {\em Soliton Equations and Hamiltonian Systems.}
Adv. Series in Math. Phys.  Vol. {\bf 12},
World Scientific, Singapore, 1991.

\bibitem{DKN}
B.A. Dubrovin, I.M. Krichever, S.P. Novikov, {\em
Integrable Systems I}. 
Enciclop\ae dya of Mathematical Sciences, vol. {\bf 4} (Dynamical Systems
IV),
V.I. Arnol'd and S.P. Novikov, eds., Springer Verlag, Berlin, 1990.

\bibitem{fmp98} G. Falqui, F. Magri, M. Pedroni, 
{\it Bihamiltonian Geometry, 
Darboux Coverings, and Linearization  of the KP Hierarchy.}
Commun.\ Math.\ Phys.\ {\bf 197} (1998), 303-324.

\bibitem{fmpz2}
G. Falqui, F. Magri, M. Pedroni, J. P. Zubelli, 
{\em A Bi-Hamiltonian theory for stationary KdV flows
and their separability.} In preparation.

\bibitem{fmt} G. Falqui, F. Magri, G. Tondo, {\em Bihamiltonian systems and
    separation  of variables: an example from the Boussinesq hierarchy.} 
solv-int/9906009, to appear in {\sl Theor. Math. Phys}.

\bibitem{GZ93} I.M. Gel'fand, I. Zakharevich,
{\em On the local geometry of a bi-Hamiltonian structure.}
In The Gel'fand Mathematical Seminars 1990-1992
(L. Corwin et al., eds.), Birkh\"auser, Boston, 1993, pp.\ 51--112.

\bibitem{GZ99}
I.M. Gel'fand, I. Zakharevich,
{\em Webs, Lenard schemes, and the local geometry of \bih\ Toda and Lax
structures.} math-ag/9903080, to appear in {\sl Selecta Mathematica}.

\bibitem{KM} Y. Kosmann--Schwarzbach, F. Magri, {\em Poisson--Nijenhuis
    structures.} Ann. Inst. Poincar\'e (Phys. Theor) {\bf 53} 
(1990), 35--81.

\bibitem{Ku81}
B.A. Kupershmidt,
{\em On the nature of the Gardner transformation.}
\jmp{22}{1981}{449--451}.

\bibitem{LM}
P. Libermann, C.M. Marle, {\em Symplectic Geometry and Analytical Mechanics}.
Reidel, Dordrecht, 1987.

\bibitem{Ma90} F. Magri, {\em Geometry and soliton equations}. In: {\sl La 
M\'ecanique Analytique de Lagrange et son h\'eritage}, Atti
Acc. Sci. Torino Suppl. {\bf 124} (1990), 181--209.

\bibitem{Pondi}
F. Magri, {\em Eight lectures on Integrable Systems.}
In: Integrability of Nonlinear Systems (Y. Kosmann-Schwarzbach et
al. eds.), Lecture Notes in Physics {\bf 495}, Springer Verlag, 
Berlin-Heidelberg, 1997, pp.\ 256--296. 

\bibitem{Magri-Magnano}
G. Magnano, F. Magri, {\em Poisson--Nijenhuis Structures and Sato Hierarchy.}
Rev.\ Math.\ Phys. {\bf 3} (1991), 403-466.

\bibitem{MR86} J.E. Marsden, T. Ratiu, {\em Reduction of Poisson Manifolds.}
\lmp{11}{1986}{161--169}.

\bibitem{Matveev-Salle} 
V.B. Matveev, M.A. Salle, {\em Darboux transformations and solitons}.
Springer Verlag, New York, 1991.

\bibitem{MGK68}
R.M. Miura, C.S. Gardner, M.D. Kruskal, {\em Korteweg-de Vries
equation and generalizations. II. Existence of conservation laws and constants
of motion.} \jmp{9}{1968}{1204--1209}.

\bibitem{Ne-B85} 
A. C. Newell, {\em Solitons in Mathematics and
Physics.} S.I.A.M., Philadelphia, 1985.

\bibitem{RSTS}
A.G. Reyman, M.A. Semenov-Tian-Shansky, {\em Compatible Poisson structures for 
Lax equations: an $r$-matrix approach.}
Phys.\ Lett.\ A {\bf 130} (1988), 456--460.

\bibitem{Sk} E. Sklyanin, {\em Separations of variables:
new trends.} Progr. Theor. Phys. Suppl. {\bf 118} (1995), 35--60.

\bibitem{SS82}
M. Sato, Y. Sato,
{\em Soliton equations as dynamical systems on infinite--dimensional 
Grassmann manifold.}
Nonlinear PDEs in Applied Sciences (US-Japan Seminar, Tokyo), 
P. Lax and H. Fujita, eds., 
North-Holland, Amsterdam, 1982, pp.\ 259--271.

\bibitem{SW85}
G. Segal, G. Wilson, {\em Loop Groups and equations of the KdV type.}
Publ. Math. IHES {\bf 61} (1985), 5--65.

\bibitem{Ta89} K. Takasaki, {\em Geometry of Universal Grassmann
Manifold from Algebraic point of view.}
Rev. Math. Phys. {\bf 1} (1989), 1--46.

\bibitem{To95} G. Tondo, {\em On the integrability of stationary and
restricted flows of the KdV hierarchy.}  J. Phys. A {\bf 28} (1995),
5097--5115.

\bibitem{Vaisman}
I. Vaisman, {\em Lectures on the Geometry of Poisson Manifolds}.
Progress in Math., Birkh\"auser, 1994.

\bibitem{Wil81}
G. Wilson, {\em On two Constructions of Conservation Laws for 
Lax equations.} Quart. J. Math. Oxford {\bf 32} (1981), 491--512. 

\end{thebibliography}
\end{document}